\def\HAL{1}

\if\HAL0
    \documentclass[AMA,LATO1COL]{WileyNJDv5}
    
    \articletype{Article Type}
    
    \received{<day> <Month>, <year>}
    \revised{<day> <Month>, <year>}
    \accepted{<day> <Month>, <year>}
    
    \journal{Journal}
    \volume{00}
    \copyyear{2023}
    \startpage{1}
\else
    \documentclass[a4paper,11pt]{article}

    \usepackage{authblk}
    
    \usepackage{fullpage}
    \usepackage{listings}
\fi

\usepackage{comment,tikz}
\usetikzlibrary{positioning,fit}

\usepackage{algorithm,algpseudocodex}

\usepackage{listings}

\usepackage{amsmath, amsthm, amsfonts,bbm}
\usepackage{stmaryrd}

\usepackage{caption}
\usepackage{subcaption}
\usepackage{url}

\theoremstyle{definition}

\theoremstyle{remark}

\usepackage{xcolor}

\newcommand{\pierre}[1]{#1} 
\newcommand{\david}[1]{#1} 

\newcommand{\bilel}[1]{#1} 
\newcommand{\todo}[1]{#1} 

\newcommand\tempnewpage[1]{ }

\begin{document}

\title{ Massively parallel CMA-ES with increasing population}

\if\HAL0
    \titlemark{ Massively parallel CMA-ES with increasing population}
\else
    
\fi

\if\HAL0

    \author[1]{David Redon}
    \author[2]{Pierre Fortin}
    \author[1]{Bilel Derbel}
    \author[3]{Miwako Tsuji}
    \author[3]{Mitsuhisa Sato}

    \authormark{David Redon, Pierre Fortin, Miwako Tsuji, Mitsuhisa Sato, Bilel Derbel}

    \address[1]{Univ. Lille, CNRS, Inria, Centrale Lille, UMR 9189 CRIStAL, F-59000 Lille, France}
    \address[2]{Univ. Lille, CNRS, Centrale Lille, UMR 9189 CRIStAL, F-59000 Lille, France}
    \address[3]{RIKEN Center for Computational Science, Kobe, Hyogo, Japan}

    \corres{\{david.redon, pierre.fortin, bilel.derbel\}\\@univ-lille.fr\\\{miwako.tsuji, msato\}@riken.jp}
    
\else

    \author[1]{David Redon}
    \author[2]{Pierre Fortin}
    \author[1]{Bilel Derbel}
    \author[3]{Miwako Tsuji}
    \author[3]{Mitsuhisa Sato}

    \affil[1]{Univ. Lille, CNRS, Inria, Centrale Lille, UMR 9189 CRIStAL, F-59000 Lille, France}
    \affil[2]{Univ. Lille, CNRS, Centrale Lille, UMR 9189 CRIStAL, F-59000 Lille, France}
    \affil[3]{RIKEN Center for Computational Science, Kobe, Hyogo, Japan
    
        \textbf{Emails}: \{david.redon, pierre.fortin, bilel.derbel\} @univ-lille.fr
        
        \{miwako.tsuji, msato\} @riken.jp
    }
\fi

\if\HAL1
    \maketitle
\fi

\if\HAL1
    \begin{abstract}
\else
    \abstract[Abstract]{
\fi


    The Increasing Population Covariance Matrix Adaptation Evolution Strategy (IPOP-CMA-ES) algorithm is a 
    reference 
    stochastic optimizer dedicated to blackbox optimization, where no prior knowledge about the  
    underlying problem structure is available. 
    This paper aims at accelerating IPOP-CMA-ES thanks to high performance computing and parallelism when solving large optimization problems. We first show how BLAS and LAPACK routines can be introduced in linear algebra operations, and we then propose two strategies for deploying 
    IPOP-CMA-ES efficiently on large-scale parallel architectures with 
    thousands of CPU cores.
    The first parallel strategy processes the multiple searches in the same ordering 
    as the sequential IPOP-CMA-ES, while the second one processes concurrently 
    these multiple searches. 
    These strategies are implemented in MPI+OpenMP and compared on 
    6144 cores of the supercomputer Fugaku. 
    We manage to obtain substantial speedups (up to several thousand) and even super-linear ones, and 
    we provide 
    an in-depth analysis of our results to understand precisely the superior performance of our second strategy. 

\if\HAL1
    \end{abstract}
\else
    }
\fi

\if\HAL1
    \def\keywords{\vspace{.5em}
    {\textit{Keywords}:\,\relax%
    }}
    \def\endkeywords{\par}
\fi


\if\HAL0
    \maketitle
\fi

\section{Introduction}

Optimization problems are prevalent in numerous modern scientific and engineering domains, 
necessitating 
increasingly intricate and compute intensive algorithms. 
They can also be of different nature depending on the application domain, the underlying computational complexity, the extent of information made available to the solvers, etc.
This paper 
addresses blackbox continuous optimization problems 
requiring large-scale parallel architectures.
More precisely, 
the goal is to find a real-valued solution $\mathbf{x}\in \mathbb{R}^n$ that minimizes (or maximizes) a continuous 
objective function $f: \mathbb{R}^n \rightarrow \mathbb{R}$. 
The function $f$ is given as blackbox,
meaning that no predetermined mathematical knowledge is available about the function, such as its derivatives or any information about its structure
(e.g. is 
the function 
smooth or convex?). 
A blackbox optimization algorithm then operates by probing $f$ for input $\mathbf{x}$, obtaining 
the corresponding fitness value $f(\mathbf{x})$, and proceeding accordingly for the rest of the search process in an iterative manner.
Blackbox optimization problems have received 
considerable attention for being essential in 
many application domains.
These include complex engineering models or numerical simulations, and generally 
application fields where problem specifics remain elusive.
For instance, applications in aeronautics necessitate optimizing aerodynamics through simulations of airflow around vehicles, 
nuclear physics involves simulating heat or particle diffusion to optimize the dimensions of containment vessels, urban transportation systems require optimizing a traffic flow determined by the patterns of stoplights, etc.
~
Traditional gradient-based approaches cannot be applied to such problems. This led to the development
of various classes of blackbox optimization algorithms, also known as derivative-free algorithms ~\cite{larson2019derivative,eiben2003introduction}.

We are interested here in the so-called CMA-ES (Covariance Matrix Adaptation Evolution Strategy) algorithm \cite{CMA}, and more specifically in 
its IPOP-CMA-ES\cite{IPOP} (Increasing Population CMA-ES) variant.
CMA-ES is a state-of-the-art blackbox optimization algorithm, 
which is, along with its variants, 
the best optimizer in the GECCO black-box optimization competition \cite{gecco_BBOB}.
CMA-ES also has applications in many fields: neural networks \cite{CMA_neural},
applied physics and engineering (e.g. for the design of thermal cloaks \cite{CMA_thermal}, optic cloaks \cite{CMA_optical}, lenses \cite{CMA_lens}, gas turbines \cite{CMA_turbine}),
autonomous sailing \cite{CMA_berthing1,CMA_berthing2}, hydrology \cite{CMA_aquifer}, sensor networks \cite{CMA_sensor}, wind energy \cite{CMA_wind}, solar energy \cite{CMA_solar}, to cite a few. 
CMA-ES is an iterative algorithm: at each iteration, it samples a set 
of $\lambda$ points (called the population) 
using a probability distribution determined by the current mean point and a $n\times n$ covariance matrix, with $n$ the dimension of the objective function $f$. The qualities of the $\lambda$ points are evaluated with $f$ and used to update the mean point 
and the matrix for the next iteration. These updates 
involve linear algebra operations 
and aim at directing the search 
towards interesting neighbouring areas.
The IPOP-CMA-ES\cite{IPOP} (Increasing Population CMA-ES) restart strategy 
improves CMA-ES, especially for more complex problems\cite{gecco_BBOB}. After one CMA-ES execution (also referred to as a {\it descent}) 
ends, because it is for example trapped 
in a local optimum, a next CMA-ES descent is started with a greater population size, 
and so on. This enables IPOP-CMA-ES to employ increasingly thorough searches, at the cost of more and more function evaluations, to eventually find better optima. 

CMA-ES and IPOP-CMA-ES can thus be used to tackle challenging blackbox  
optimization problems.  
However, large optimization problems still require a lot of compute power, 
because the function evaluations can be individually time-consuming, 
and/or because 
many function evaluations (i.e. CMA-ES iterations) can be needed to solve complex problems. 
Two main approaches can be distinguished in the literature to accelerate CMA-ES and IPOP-CMA-ES for large optimization problems.
The first approach focuses on reducing the cost of the linear algebra operations 
while retaining as much quality as possible in the search for 
solutions.
This is done by storing a reduced number of parameters instead of a full matrix of $n^2$ parameters,
leading to reduced costs for updating 
and using the matrix. 
These so-called large-scale CMA-ES variants 
can store 
$k<n$ vectors of $n$ parameters \cite{Rm_CMA_ES,LM_MA_ES,VkD_CMA_ES,LM_CMA_ES,Fast_CMA,CMA_MMES}, 
or only $n$ parameters \cite{sep_CMA_ES},
or even only 
the sampled points\cite{CMA_gradients}. 
Subsequent works compared the respective merits of some of those large-scale variants\cite{CMA_large_scale,CMA_large_scale_global,CMA_large_scale_bfgs} 
taking into account 
their lower 
ability to retain information about the local function landscape, which leads to less effective convergence and to a lower quality for the final solution.
The second approach does not degrade the solution quality but relies on parallelism 
for the independent function evaluations
\cite{pCMALib,CMA_auto_adapt}.
This 
was for example used for  
specific application domains
\cite{CMA_neural,CMA_flow,CMA_traffic}.
Additionally, 
certain parameters can be adapted during the CMA-ES descent\cite{CMA_auto_adapt}, 
in the hope of finding settings which best fit the parallel hardware. 
Another way of using parallelism 
is to run multiple CMA-ES descents concurrently, 
while trying to improve 
the convergence of any given descent using information from the other descents.
In optimization, this method is known as the island model.
Some island models were proposed for the original CMA-ES\cite{PSCMAES,pCMALib,CMAES_island_async}.
Such island models have also been used in specific domains\cite{CMA_reacts,CMA_islands_funnels,CMA_calibr}, or for 
multi-objective optimization\cite{CMA_islands_multiobj}.
Some works also implement an island model for a large-scale CMA-ES variant \cite{CMA_async_limited_memory,D_LM_MA,CMA_DES}.
However, to the best of our knowledge, there currently exists no work on the parallelization of IPOP-CMA-ES, which implies running descents of increasing 
population sizes. 
~

In this paper, we thus focus on the use of parallelism and HPC to solve large optimization problems with IPOP-CMA-ES by speeding up both the linear algebra operations and the function evaluations. 
We present the following contributions.

\begin{itemize}
    \item
    We first show how the CMA-ES linear algebra operations can be accelerated thanks to BLAS and LAPACK routines. This requires the rewrite of some of these operations in order to introduce more efficient BLAS routines. 
    \item 
    We present two parallel strategies for IPOP-CMA-ES to fully exploit a large number of CPU cores (up to several thousand). Such a number of CPU cores implies multiple compute nodes in distributed memory, each node being composed of multiple cores in shared memory. The goal here is to leverage large-scale parallelism (via multiple nodes) 
    to benefit from the increasing 
    number of (parallel) evaluations in IPOP-CMA-ES.
    The first strategy performs descents in the same order of population size as the original IPOP-CMA-ES, while the second strategy processes concurrently 
    descents of different population sizes. 
    \item
    We thoroughly compare MPI\cite{mpi}+OpenMP\cite{openmp} 
    implementations of our two strategies on 6144 cores (128 nodes) of the supercomputer Fugaku in order to determine which one is the most relevant on such a large-scale parallel architecture.  
    These comparisons are performed with the reference BBOB\cite{Hansen10real-parameterblack-box} (Black-Box Optimization Benchmarking) benchmark for various dimensions and various function evaluation costs. 
    We also present a fine analysis of those results, aggregated in different ways to investigate the impacts of the features of the parallel strategies on performance and on quality, depending on the targeted function, on the dimensions and on the evaluation costs. 
\end{itemize}

The rest of this paper is organized as follows.
In Section \ref{section:CMAES}, we give some background on CMA-ES and on IPOP-CMA-ES.
In Section \ref{section:parallelDesigns}, we show how we have introduced BLAS and LAPACK routines, and we describe the proposed parallel strategies.
We report our detailed experimental study in Section \ref{section:results}, and we conclude the paper in Section \ref{section:conclusion}.

\section{CMA‐ES with increasing population}
\label{section:CMAES}

We discuss here the main working principles of the Covariance Matrix Adaptation Evolution Strategy with Increasing Population\cite{IPOP} (IPOP-CMA-ES), starting with the Covariance Matrix Adaptation Evolution Strategy (CMA-ES) it is based on.

\subsection{The Covariance Matrix Adaptation Evolution Strategy}
\label{s:pres_cmaes}

\begin{figure}[t] 
    \centering
    \includegraphics[width=\linewidth,clip]{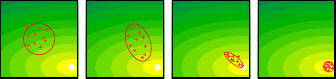}
    \caption{Convergence example of CMA-ES on a function space.
    The white dot indicates the function optimum, 
    the red ellipse the normal law, and the red crosses points sampled according to this law.
    }
    \label{fig:cmaes_converg}
\end{figure}

\begin{algorithm}[t] 
\caption{Pseudocode of CMA-ES with population size $\lambda$, for an objective function $f$}
\label{algo:cmaes}
\begin{algorithmic}[1]
    \State initialize (mean: $m$, covariance matrix: $C$, variance: $\sigma$, evolution path of $\sigma$: $p_{\sigma}$, evolution path of $C$: $p_c$)
    \While {no stopping criterion met}
        \LComment{Perform one CMA-ES iteration}
        \For {$i = 1..\lambda$}\label{CMA_ES_start}\label{a:sampling_start} %
            \State $x_i \gets$ sample\_point$(m,C,\sigma)$
        \EndFor\label{a:sampling_end}
        \For {$i = 1..\lambda$}
            \State $f_i \gets f(x_i)$
        \EndFor
        \State update $(m,C,\sigma,p_{\sigma},p_c)$ using $(x_i)_{i=1..\lambda}$ and $(f_i)_{i=1..\lambda}$ \label{a:update}\label{CMA_ES_end}
    \EndWhile
    \State \Return the sampled point with the best quality 
\end{algorithmic}
\end{algorithm}

In order to find the optimum of an objective function $f$ of dimension $n$, CMA-ES maintains a best current point (also called the $mean$ or $m$) in $\mathbb{R}^n$. At each iteration, CMA-ES samples $\lambda$ points in $\mathbb{R}^n$ ($\lambda$ is called the {\it population size}) around its mean $m$ according to a normal law distorted along an ellipsoid \cite{MatCov}: see Figure~\ref{fig:cmaes_converg}. 
This normal law samples points in a circular fashion around the mean, using more points near the mean, and fewer points away from it. 
The widths and orientations of the ellipsoid are given by the so-called {\it covariance matrix} $C$. 
That is, sampling points involves a multivariate normal distribution $\mathcal{N}(0,C)$ with zero mean and covariance matrix $C$. This matrix is $n\times n$ shaped: CMA-ES can thus adapt the space in which it samples points to the local shape of the function.

The mean $m$, as well as the matrix $C$, are updated depending on the qualities of the sampled points. These qualities correspond to the evaluations of the function $f$ at all sampled points. 
For the covariance matrix, this update is called a matrix adaptation. This modification of the matrix aims at distorting the ellipsoid in ways that make it more likely to sample new points in the positions (relative to $m$) the best previous points were found in.
This update requires $O(\lambda \times n^2)$ operations.

The scale at which CMA-ES searches for new points is also determined by the variance $\sigma$ of the normal law. This variance $\sigma$ is updated 
using a so-called {\it evolution path} $p_\sigma$ \cite{cumul}. The evolution path is the sum of the last shifts of $m$. If $m$ is generally shifting in one direction (the better points being farther along a certain axis), then the variance is increased. If $m$ is not shifting in a specific direction (the ellipse remains in the same area, the better points being around $m$), then the variance is decreased. A similar technique is used for the matrix adaptation, with an evolution path $p_c$.

In the end, the return value of CMA-ES is the sampled point with the best quality. 
A high-level algorithm of CMA-ES is given in Algorithm~\ref{algo:cmaes}, and its convergence is illustrated in Figure~\ref{fig:cmaes_converg}.

\medskip 
We now provide a few more mathematical details. 
Sampling from $\mathcal{N}(0,C)$ requires the matrices $B$ and $D$, where 
$B$ is a matrix containing orthonormal eigenvectors of $C$,
and $D$ is a diagonal matrix containing the square roots of the eigenvalues of $C$.
Eventually, we want to sample the points $x_k \in \mathbb{R}^n$ from the normal law $\mathcal{N}(m,\sigma^2 C)$. The corresponding equation reads
\begin{equation} \label{eq:1}
x_k = m + \sigma B D z_k, \quad \forall 1 \le k \le \lambda
\end{equation}
with $z_k \in \mathbb{R}^n$ sampled from $\mathcal{N}(0,I)$, $I$ being the $n \times n$ identity matrix. 
~
The equation for adapting the covariance matrix $C$ according to the qualities of the newly sampled points then reads
\begin{equation} \label{eq:2}
C \leftarrow C + c_\mu \sum_{i=1}^{\lambda}{ w_{rk(i)} (y_i y_i^T - C) } + c_1(p_c p_c^T - C)
\end{equation}
where $c_\mu \in \mathbb{R}$ and $c_1 \in \mathbb{R}$ are learning rate parameters for CMA-ES,
$w_{rk(i)} \in \mathbb{R}$ is a weight depending on the rank of the point $x_i$ when sorted by function value (points with better function values having greater weights, and $\sum_{i=1}^{\lambda}{ w_{rk(i)} } = 1$),
and $\forall i \in \llbracket 1, \lambda \rrbracket, y_i \in  \mathbb{R}^n$ is such that $x_i = m + \sigma y_i$ \cite{CMA_rangMu}.
~
In turn, computing $B$ and $D$ from the matrix $C$ involves an eigendecomposition, which requires $O(n^3)$ operations with $n$ the function dimension.
Updating other variables like $p_c$ or $p_\sigma$ 
only requires at most $O(n^2)$ or $O(\lambda \times n)$ operations and is thus less time-consuming.

\subsection{The increasing population restart strategy} 

\begin{algorithm}[t] 
\caption{Pseudocode of IPOP-CMA-ES with initial population size $\lambda_{start}$,  multiplicative factor $2$, maximum coefficient $K_{max}$, for an objective function $f$}
\label{algo:ipop}
\begin{algorithmic}[1]
    \State $K \gets 1$
    \While {$K \leq K_{max}$ {\bf and} \emph{budget not exhausted}}
        \State initialize (mean: $m$, covariance matrix: $C$, variance: $\sigma$, evolution path of $\sigma$: $p_{\sigma}$, evolution path of $C$: $p_c$) 
        \While {no stopping criterion met}
            \State process one iteration of CMA-ES (see Algorithm \ref{algo:cmaes}, lines \ref{CMA_ES_start}-\ref{CMA_ES_end}) using $m$, $C$, $\sigma$, $p_{\sigma}$, $p_c$ and with population size $K \times \lambda_{start}$
        \EndWhile
        \State $K \gets K \times 2$
    \EndWhile
    \State \Return the sampled point with the best quality (over all descents) 
    \end{algorithmic}
\end{algorithm}

CMA-ES is a stochastic algorithm due to the random sampling of points within a distribution. Two executions of Algorithm \ref{algo:cmaes} (also referred to as {\it descents}) on the same problem with the same parameters may thus sample different points and return a different best point.
One execution may indeed find high quality points the other missed. As such, executing CMA-ES multiple times enables one to increase the quality of the final best point. 
The exact benefit of these multiple executions will of course depend on the shape of the objective function.
In order to identify if the search has settled on a local optimum and if the current execution must stop, multiple stopping conditions\cite{LR} have been proposed for CMA-ES.
The stopping conditions can generally be understood as either the function quality not improving anymore, or the function being locally too flat, or even the sampling distribution being too small (i.e. the mean stops moving, or almost so).
The best is then to restart the algorithm (i.e. start a new descent) in the hope of finding even better solution points.

It has then been shown that increasing the population size $\lambda$ for each new restart enables faster convergence \cite{IPOP}.
More precisely, this CMA-ES Increasing Population restart scheme (IPOP-CMA-ES) offers the same convergence rate as CMA-ES on simple functions, and a significantly better one on many complex functions.
The corresponding high-level algorithm is presented in Algorithm~\ref{algo:ipop}.
As usual with IPOP-CMA-ES \cite{IPOP,ipop_factor2_ref1,ipop_factor2_ref2,ipop_factor2_ref3}, 
we rely on a multiplicative factor of 2 at each restart 
for the population size. The initial population size $\lambda_{start}$ is hence multiplied by $K=2^i$ 
for the $i$-th CMA-ES execution. Hence $K$ ranges from $1$ to a certain $K_{max}$, with for example $K_{max}=2^8$. 
~
At each iteration multiple function evaluations are performed and the number of evaluations equals the population size.
Since these evaluations are embarrassingly parallel, IPOP-CMA-ES offers thus an increasing degree of parallelism along its execution. 
When targeting a parallel version of IPOP-CMA-ES, this increasing degree of parallelism is an opportunity to reach important parallel speedups, but one  also requires a relevant strategy to deploy at best this varying parallelism degree on a given (fixed) number of CPU cores. We will thus study different parallel strategies in the next section.

\section{High performance parallel strategies}
\label{section:parallelDesigns}

\subsection{High performance linear algebra}
\label{s:blas_lapack}

Before targeting large-scale parallel speedups via different parallel strategies, we first consider the performance of IPOP-CMA-ES and we study here how BLAS and LAPACK routines can be introduced in CMA-ES (hence in IPOP-CMA-ES). 
BLAS (Basic Linear Algebra Subprograms\footnote{See: \url{https://www.netlib.org/blas/}}) are high-performance implementations of standard linear algebra operations: vector operations (Level 1 BLAS), matrix-vector operations (Level 2 BLAS), and matrix-matrix operations (Level 3 BLAS). LAPACK (Linear Algebra PACKage\footnote{See: \url{https://www.netlib.org/lapack/}}) then relies on BLAS routines to efficiently solve e.g. systems of linear equations, eigenvalue problems, singular value problems, etc.   
As presented in Section \ref{s:pres_cmaes}, multiple steps in CMA-ES actually involve 
linear algebra operations. 
We focus on Level 3 BLAS operations with a cubic time complexity for a quadratic input data size: these provide indeed a linear arithmetic intensity which enables their implementation to highly take advantage of current CPU architectures. 
We have hence been able to introduce BLAS and LAPACK routines in the following three steps, either straightforwardly or thanks to some rewriting of the linear algebra operations. 

\newcommand*{\vrtibar}{\rule[-1ex]{0.5pt}{2.5ex}}
\newcommand*{\horibar}{\rule[.5ex]{2.5ex}{0.5pt}}
\newcommand{\vectormatrix}[3]{
\begin{pmatrix}
    \vrtibar && \vrtibar && \vrtibar \\
    #1 & \cdots & #2 & \cdots & #3 \\
    \vrtibar && \vrtibar && \vrtibar
\end{pmatrix}}

\begin{itemize}
    \item The eigendecomposition of the covariance matrix $C$ can first benefit easily from LAPACK by using the $dsyev$ routine. 

\item Second, the original equation (see equation \ref{eq:2}) for the covariance matrix adaptation does not involve any matrix-matrix multiplication. 
However, we can first rewrite equation \ref{eq:2} as
\begin{equation*}
C \leftarrow (1 - c_\mu - c_1) C + c_\mu  ( \sum_{i=1}^{\lambda}{ w_{rk(i)} y_i y_i^T}) + c_1 p_c p_c^T, \textrm{~since~} \sum_{i=1}^{\lambda}{ w_{rk(i)} } = 1.
\end{equation*}
We denote by $M$ the $n \times n$ matrix equal to $\sum_{i=1}^{\lambda}{w_{rk(i)} y_i y_i^T}$.
We have then
\begin{equation*} 
\forall (r,c) \in \llbracket 1, n \rrbracket^2, M_{r,c} =  \sum_{i=1}^{\lambda} { w_{rk(i)} (y_i)_r (y_i^T)_c} = \sum_{i=1}^{\lambda} { (y_i)_r (w_{rk(i)}  (y_i^T)_c)} = \sum_{i=1}^{\lambda} { A_{r,i} B_{i,c} }  
\end{equation*}
where $A$ is a $n \times \lambda$ matrix containing columns of $(y_i)_{i=1..\lambda}$ and $B$ is a $\lambda \times n$ matrix containing rows of $(w_{rk(i)} y_i^T)_{i=1..\lambda}$, namely: 
\begin{center} $A = \vectormatrix{y_1}{y_k} {y_{\lambda}}$ , $B = \begin{pmatrix}
    \horibar & w_{rk(1)} y_1^T & \horibar \\
     & \vdots & \\
    \horibar & w_{rk(k)} y_k^T & \horibar \\
     & \vdots & \\
    \horibar & w_{rk(\lambda)} y_{\lambda}^T & \horibar
\end{pmatrix}$. \end{center} 

In the end, we have rewritten the covariance matrix adaptation as 
\begin{equation} \label{eq:3}
C \leftarrow (1 - c_\mu - c_1) C + c_\mu A \cdot B + c_1 p_c p_c^T
\end{equation}

where the matrix product $A \cdot B$ can be very efficiently performed thanks to the Level 3 $dgemm$ BLAS.
$2 \lambda n$ affectations are required to create the matrices $A$ and $B$, but this cost will be dominated by the $\lambda n^2$ operation cost of the matrix product: the BLAS performance gain is thus likely to (at least partly) offset these extra affectations.  
Moreover, the sizes of matrices $A$ and $B$ depend on $\lambda$: this makes our BLAS rewriting relevant for IPOP-CMA-ES for two reasons. First, the 
covariance matrix adaptation is more time-consuming for IPOP-CMA-ES, where the population size $\lambda=K \lambda_{start}$ is increasing with $K$ 
and becomes eventually larger than for a standard CMA-ES execution. 
Second, since best $dgemm$ performance is obtained for large enough matrices, the BLAS gain will thus be stronger for IPOP-CMA-ES. 
Finally, it can be noticed that this specific covariance matrix adaptation was already used in the Fortran source code of the CMA-ES {\em pCMALib} library, but such rewriting, and its performance impact (especially for IPOP-CMA-ES), were not presented in the corresponding paper\cite{pCMALib}.


\item Finally, regarding the sampling step, equation \ref{eq:1} implies only matrix-vector products (since $D$ is a diagonal matrix). We can however compute all $(x_k)_{k=1..\lambda}$ at once using:
\[ \vectormatrix{x_1}{x_k}{x_{\lambda}} =
\vectormatrix{m}{m}{m} + \sigma B D \cdot \vectormatrix{z_1}{z_k}{z_{\lambda}}. \]
In practice, this only implies $\lambda n$ extra affectations
to fill a $n \times \lambda$ matrix with the mean vector $m$. Again, this extra cost will be dominated by the matrix mutliplication cost ($\lambda n^2$ operations), and likely be offset by the BLAS performance gain. 
Moreover, as for the covariance matrix adaptation, the matrix sizes depend on $\lambda$ which makes this step more time-consuming and the BLAS performance gain stronger for IPOP-CMA-ES than for CMA-ES.
To our knowledge, no so such rewrite of the CMA-ES sampling equation was proposed before.

\end{itemize}

\subsection{The parallel strategies}

In this section, we aim at deploying IPOP-CMA-ES on large-scale parallel architectures with thousands of CPU cores. 
Such architectures are based on muliple nodes (distributed-memory parallelism) composed on few 
multi-core processors each (shared-memory parallelism): we will thus rely on a hybrid MPI+OpenMP programming, using MPI for inter-process communications and OpenMP 
for multi-thread parallelism (with $T$ threads in each MPI process). We consider an IPOP-CMA-ES execution with population sizes $K \times \lambda_{start}$, $K$ being a power of $2$ ranging from $2^0$ to $K_{max}$ (see Algorithm \ref{algo:ipop}). For such an IPOP-CMA-ES execution, we propose here two 
generic strategies to obtain the best parallel speedups on a fixed (arbitrarily large) number of CPU cores.  
These strategies will then be specified and compared on a given supercomputer (Fugaku) in Section \ref{section:results}.

\subsubsection{Parallelism within a CMA-ES descent}
\label{s:parallel_descent}

First of all, in our parallel strategies, each CMA-ES descent will be driven by a dedicated MPI process, hereafter referenced to as the {\it main} process.  
At each iteration of a descent, CMA-ES evaluates the objective function on $\lambda$ points: these $\lambda$ evaluations can be performed in parallel.
In order to obtain the best parallel speedups, we aim at fully exploiting this parallelism level by processing each evaluation on a dedicated CPU core.  

\begin{itemize}
    \item When $\lambda \leq T$, we can distribute the $\lambda$ evaluations on the $T$ threads of the main process. 

\item When $\lambda > T$, we have to rely on multiple MPI processes. The main process will thus 
first generate the list of points where the objective function must be evaluated. These points are then "scattered" (using the corresponding MPI function) on a set of MPI processes. This sets an implicit synchronization among all processes involved in the same descent. Each of these MPI processes will evaluate the function on its points (using its $T$ threads), and the objective function values are finally "gathered" (using again the corresponding MPI function) back to the main process. 
\end{itemize}

Finally, we will also consider multi-thread parallelism for the linear algebra operations performed in each descent (see Section \ref{s:blas_lapack}): this will be detailed in Section \ref{sectionLinalgResults}.

\subsubsection{The K-Replicated strategy}

\if\HAL0
    \newcommand{\siHorsTrou}[2]{
        \pgfmathparse{#1<\debutTrou || #1>\finTrou ?int(1):int(0)}
        \ifnum \pgfmathresult>0
            #2
        \fi
    }

    \newcommand{\IPOPparalNaifAuxATrou}[3]{%
        \ifnum #3>0
            \pgfmathsetmacro \position{#2*(#3-1)}
            
            \siHorsTrou{#3}{
                \draw[fill=gray!20] (\position,0) rectangle ({\position+#2},{#1});
            }
            
            \pgfmathsetmacro\newheight{#1/2}
            \IPOPparalNaifAuxATrou{\newheight}{#2}{\numexpr#3-1\relax}
        \fi
    }
    \newcommand{\IPOPparalNaifATrou}{
        
        \pgfmathsetmacro \hauteur{5}
        \pgfmathsetmacro \largeur{5}
        \pgfmathsetmacro \nRectangles{9}
        \pgfmathsetmacro \debutTrou{5}
        \pgfmathsetmacro \finTrou{6}
        \pgfmathsetmacro \largeurDesc{\largeur / \nRectangles}
        \IPOPparalNaifAuxATrou{\hauteur}{\largeurDesc}{\nRectangles}
        
        \foreach \i in {0,...,\numexpr\nRectangles-1} {
            \ifnum \i<4
                \node at ({(\i+0.5) * \largeurDesc},-3mm) {$2^{\i}$};
            \fi
            \ifnum \i>5
                \ifnum \i<7
                    \node at ({(\i+0.5) * \largeurDesc},-3mm) {$\frac{K_{max}}{2^2}$};
                \else \ifnum \i<8
                    \node at ({(\i+0.5) * \largeurDesc},-3mm) {$\frac{K_{max}}{2}$};
                \else
                    \node at ({(\i+0.5) * \largeurDesc},-3mm) {$K_{max}$};
                \fi \fi
            \fi
        }

        \node at (2.75,3mm) {\LARGE $\dots$};
        \node at (2.75,-3mm) {\LARGE $\dots$};

        \draw[->] (0,-6mm) -- (\largeur,-6mm) node[midway, below] {time / descent with $K=2^i$};
        \draw[->] (-6mm,0) -- (-6mm,\hauteur) node[midway, right] {core occupancy};
    }

    \newcommand{\IPOPparalSaturAuxAtrou}[7]{%
        \ifnum #3>#6
            \pgfmathsetmacro \position{#2*(#5+#3-1)}
            
            \ifnum #3<#7
                \draw[fill=gray!20, line width = 0.4pt] (\position,#4) rectangle ({\position+#2},{#1+#4});
            \fi

            \IPOPparalSaturAuxAtrou{#1/2}{#2}{\numexpr#3-1\relax}{#4}{#5}{#6}{#7}
            \IPOPparalSaturAuxAtrou{#1/2}{#2}{\numexpr#3-1\relax}{#4+#1/2}{#5}{#6}{#7}
        \fi
    }
    \newcommand{\IPOPparalSaturATrou}{
        \pgfmathsetmacro \hauteur{5}
        \pgfmathsetmacro \largeur{5}
        \pgfmathsetmacro \nRectangles{8}
        \pgfmathsetmacro \largeurDesc{\largeur / \nRectangles}

        \IPOPparalSaturAuxAtrou{\hauteur}{\largeurDesc}{\nRectangles}{0} {-1}{1}{5}
        \IPOPparalSaturAuxAtrou{\hauteur}{\largeurDesc}{\nRectangles}{0} {0}{5}{9}
        
        \pgfmathsetmacro \debutTrou{4}
        \pgfmathsetmacro \finTrou{5}
        \foreach \i in {0,...,\numexpr\nRectangles-1} {
            \ifnum \i<3
                \node at ({(\i+0.5) * \largeurDesc},-3mm) {$2^{\i}$};
            \fi
            \ifnum \i>4
                \ifnum \i<6
                    \node at ({(\i+0.5) * \largeurDesc},-3mm) {$\frac{K_{max}}{2^2}$};
                \else \ifnum \i<7
                    \node at ({(\i+0.5) * \largeurDesc},-3mm) {$\frac{K_{max}}{2}$};
                \else
                    \node at ({(\i+0.5) * \largeurDesc},-3mm) {$K_{max}$};
                \fi \fi
            \fi
        }
        
        \foreach \i in {0,...,3} {
            \node at (2.5, {0.625+ 1.25*\i} ) {\LARGE $\dots$};
        }
        \node at (2.5,-3mm) {\LARGE $\dots$};
        
        \draw[->] (0,-6mm) -- (\largeur,-6mm) node[midway, below] {time / descent with $K=2^i$};
        \draw[->] (-6mm,0) -- (-6mm,\hauteur) node[midway, right] {}; 
    }

    \begin{figure}
    \centering
    \begin{minipage}{.45\textwidth}
        \centering
            \scalebox{1.35} {
            \begin{tikzpicture}
                \IPOPparalNaifATrou
            \end{tikzpicture} }
        \caption{Illustration of the core occupancy of a naive version of IPOP-CMA-ES with successive parallel descents.}
        \label{fig:krepl1}
    \end{minipage}
    \hspace{.05\textwidth}
    \begin{minipage}{.45\textwidth}
        \centering
            \scalebox{1.35} {
            \begin{tikzpicture}
                \IPOPparalSaturATrou
            \end{tikzpicture} }
        \caption{Illustration of the core occupancy of the K\nobreakdashes-Replicated strategy.}
        \label{fig:krepl2}
    \end{minipage}
    \end{figure}

\else

    \newcommand{\siHorsTrou}[2]{
        \pgfmathparse{#1<\debutTrou || #1>\finTrou ?int(1):int(0)}
        \ifnum \pgfmathresult>0
            #2
        \fi
    }

    \newcommand{\IPOPparalNaifAuxATrou}[3]{%
        \ifnum #3>0
            \pgfmathsetmacro \position{#2*(#3-1)}
            
            \siHorsTrou{#3}{
                \draw[fill=gray!20] (\position,0) rectangle ({\position+#2},{#1});
            }
            
            \pgfmathsetmacro\newheight{#1/2}
            \IPOPparalNaifAuxATrou{\newheight}{#2}{\numexpr#3-1\relax}
        \fi
    }
    \newcommand{\IPOPparalNaifATrou}{
        
        \pgfmathsetmacro \hauteur{2.5mm}
        \pgfmathsetmacro \largeur{2.5mm}
        \pgfmathsetmacro \nRectangles{9}
        \pgfmathsetmacro \debutTrou{5}
        \pgfmathsetmacro \finTrou{6}
        \pgfmathsetmacro \largeurDesc{\largeur / \nRectangles}
        \IPOPparalNaifAuxATrou{\hauteur}{\largeurDesc}{\nRectangles}
        
        \foreach \i in {0,...,\numexpr\nRectangles-1} {
            \ifnum \i<4
                \node at ({(\i+0.5) * \largeurDesc},-3mm) {$2^{\i}$};
            \fi
            \ifnum \i>5
                \ifnum \i<7
                    \node at ({(\i+0.5) * \largeurDesc},-3mm) {\small $\frac{K_{max}}{2^2}$};
                \else \ifnum \i<8
                    \node at ({(\i+0.5) * \largeurDesc},-3mm) {\small $\frac{K_{max}}{2}$};
                \else
                    \node at ({(\i+0.5) * \largeurDesc},-3mm) {\small $K_{max}$};
                \fi \fi
            \fi
        }
        
        \pgfmathsetmacro \milieuTrou{ (\debutTrou+\finTrou-1)/2/\nRectangles) * \largeur }
        \node at ( {\milieuTrou} ,3mm) {\LARGE $\dots$};
        \node at ( {\milieuTrou} ,-3mm) {\LARGE $\dots$};

        \draw[->] (0,-6mm) -- (\largeur,-6mm) node[midway, below] {time / descent with $K=2^i$};
        \draw[->] (-6mm,0) -- (-6mm,\hauteur) node[midway, right] {core occupancy};
    }

    \newcommand{\IPOPparalSaturAuxAtrou}[7]{%
        \ifnum #3>#6
            \pgfmathsetmacro \position{#2*(#5+#3-1)}
            
            \ifnum #3<#7
                \draw[fill=gray!20, line width = 0.4pt] (\position,#4) rectangle ({\position+#2},{#1+#4});
            \fi

            \IPOPparalSaturAuxAtrou{#1/2}{#2}{\numexpr#3-1\relax}{#4}{#5}{#6}{#7}
            \IPOPparalSaturAuxAtrou{#1/2}{#2}{\numexpr#3-1\relax}{#4+#1/2}{#5}{#6}{#7}
        \fi
    }
    \newcommand{\IPOPparalSaturATrou}{
        
        \pgfmathsetmacro \hauteur{2.5mm}
        \pgfmathsetmacro \largeur{2.5mm}
        \pgfmathsetmacro \nRectangles{8}
        \pgfmathsetmacro \largeurDesc{\largeur / \nRectangles}

        \IPOPparalSaturAuxAtrou{\hauteur}{\largeurDesc}{\nRectangles}{0} {-1}{1}{5}
        \IPOPparalSaturAuxAtrou{\hauteur}{\largeurDesc}{\nRectangles}{0} {0}{5}{9}
        
        \pgfmathsetmacro \debutTrou{4}
        \pgfmathsetmacro \finTrou{5}
        \foreach \i in {0,...,\numexpr\nRectangles-1} {
            \ifnum \i<3
                \node at ({(\i+0.5) * \largeurDesc},-3mm) {$2^{\i}$};
            \fi
            \ifnum \i>4
                \ifnum \i<6
                    \node at ({(\i+0.5) * \largeurDesc},-3mm) {$\frac{K_{max}}{2^2}$};
                \else \ifnum \i<7
                    \node at ({(\i+0.5) * \largeurDesc},-3mm) {$\frac{K_{max}}{2}$};
                \else
                    \node at ({(\i+0.5) * \largeurDesc},-3mm) {$K_{max}$};
                \fi \fi
            \fi
        }
        
        \pgfmathsetmacro \milieuTrou{ (\debutTrou+\finTrou-1)/2/\nRectangles) * \largeur }
        \foreach \i in {0,...,3} {
            \node at ( {\milieuTrou}, {0.625+ 1.25*\i} ) {\LARGE $\dots$};
        }
        \node at ({\milieuTrou},-3mm) {\LARGE $\dots$};
        
        \draw[->] (0,-6mm) -- (\largeur,-6mm) node[midway, below] {time / descent with $K=2^i$};
        \draw[->] (-6mm,0) -- (-6mm,\hauteur) node[midway, right] {}; 
    }

    \begin{figure}
    \centering
    \begin{minipage}{.45\textwidth}
        \centering
            \scalebox{0.95} {
            \begin{tikzpicture}
                \IPOPparalNaifATrou
            \end{tikzpicture} }
        \caption{Illustration of the core occupancy of a naive version of IPOP-CMA-ES with successive parallel descents.}
        \label{fig:krepl1}
    \end{minipage}
    \hspace{.05\textwidth}
    \begin{minipage}{.45\textwidth}
        \centering
            \scalebox{0.95} {
            \begin{tikzpicture}
                \IPOPparalSaturATrou
            \end{tikzpicture} }
        \caption{Illustration of the core occupancy of the K\nobreakdashes-Replicated strategy.}
        \label{fig:krepl2}
    \end{minipage}
    \end{figure}

\fi

When targeting a parallel IPOP-CMA-ES execution, the first idea that may come to mind is 
running successively each descent of increasing $K$ value, using the parallelism available within each descent (as described in Section \ref{s:parallel_descent}). 
This is illustrated in Figure \ref{fig:krepl1}.
The downside of this approach is an overall low CPU core occupancy: since descents with large $K$ 
have a higher degree of parallelism, most of the CPU cores will be unused for the other descents.


A first solution, hereafter referred to as `K-Replicated`, is to replicate 
the current $K$ descent 
unto multiple, independent descents (with the same $K$ value) 
until all the computing resources are used: 
see Figure \ref{fig:krepl2}.
The number of those replicated descents is $c/(K \times \lambda_{start})$, where $c$ is the number of CPU cores available.
We thus have more simultaneous descents at the start, when $K$ is small, and fewer 
descents when $K$ is larger; but the overall resource usage remains the same at any time. 
This way, all CPU cores 
are being used at all time, and since CMA-ES is a stochastic algorithm, the additional descents can help finding better solutions. 

\begin{algorithm}[t] 
\caption{Pseudocode of the K-Replicated strategy. 
The global communicator \lstinline{MPI_COMM_WORLD} (containing all MPI processes) and $K{max}$ are used for the initial call.} 
\label{algo:krepl}
\begin{algorithmic}
    \Procedure{K-Replicated}{communicator, $K$}
        \If {$K > 1$}
            \State my\_rank $\gets$ \Call{MPI\_Comm\_rank}{communicator}
            \State size $\gets$ \Call{MPI\_Comm\_size}{communicator}
            \State my\_half\_comm $\gets$ \Call{MPI\_Comm\_Split}{communicator, my\_rank $\leq$ size $/2$, my\_rank}\Comment{splits `communicator` in two halves of equal size}
            \State \Call{K-Replicated}{my\_half\_comm, $K / 2$}
        \EndIf
        \If {$K \leq K_{max}$}
            \State \Call{CMA-ES descent}{communicator, $K \times \lambda_{start}$} 
        \EndIf
    \EndProcedure
\end{algorithmic}
\end{algorithm}

Regarding the implementation, once two $K$ descents 
are finished, their resources can be used for a subsequent $2 K$ descent. 
This can be efficiently implemented as in the recursive Algorithm \ref{algo:krepl} using a hierarchy of MPI communicators, which represent sets of MPI processes that can exchange 
messages \cite{mpi}. 
New communicators can be created by specifying 
subsets out of a previously existing communicator: each process of the parent communicator indicate which child communicator it belongs to. 
The global communicator is thus split until the resulting communicators are small enough to
be used for descents with $K = 1$.
Then, each time a pair of $K=2^i$ descents from a same parent communicator is finished,
the control flows back up 
to this parent communicator for a $K=2^{i+1}$ descent.
This is repeated until $K_{max}$ is reached and finished as well.
Finally, in order to have a distinct random generator seed in each CMA-ES descent, we rely on the current time multiplied by the rank of the MPI process in the global communicator. 


\subsubsection{The K-Distributed strategy}

The K-replicated strategy exploits all the available CPU cores by replicating multiple descents with the same $K$ value. These concurrent $K$ descents increase the chances of finding better solutions for the stochastic IPOP-CMA-ES algorithm. 
We now consider a second strategy to leverage parallelism among multiple CMA-ES descents 
for 
large-scale parallel architectures, 
which stems from three observations.

Firstly, a descent of population size $\lambda$ can be sped-up up to a factor of $\lambda$ (provided the communication and linear algebra times are short enough).
~
Secondly, a descent of population $K \times \lambda_{start}$ usually takes, roughly, $K$ times as long to reach a given quality as a $\lambda_{start}$ descent.
This can be interpreted as CMA-ES taking in more information about a local landscape of the function before making the choice of where to move next, which requires 
more time.
The benefit of using a larger population size is that, in many cases, the decision will be more 'informed', causing CMA-ES to avoid being trapped in a local optimum for longer and ultimately finding better solutions before the end of the descent.
~
Finally, although the previous observation \pierre{apply in general,} 
there are also \pierre{cases where,} 
for some quality ranges 
on some objective functions and for some $K$ values, 
a $K \times \lambda_{start}$ descent will be faster or slower than $K$ times the duration of a $\lambda_{start}$ descent.
We interpret these \pierre{different behaviors} 
as \pierre{the impact of the properties of the function landscape on the optimal population size}.
Note that, especially for more complex functions, the landscape properties can change depending on the region of the search space or on the scale it is observed at.

From the first and second observation, we can reasonably expect that several descents, each with a different population size $\lambda$ 
and each running on $\lambda$ CPU cores, 
will generally improve the quality of their current best solution \pierre{for roughly the same amount of computation time}. 
Then, from the third observation, we can expect that, thanks to their 
different population sizes, some of these parallel descents will be faster than others to reach certain qualities.
Therefore, running a range of population sizes in parallel may 
have better results than running several descents of the same population size, as K-Replicated does.


\newcommand{\IPOPdiffKAuxAtrou}[6]{
    \siHorsTrou {#5} {
        \draw[fill=gray!20] (#1,0) rectangle (#1+#2,#4);
        \pgfmathsetmacro \ik{int(#6)}
        \ifnum #5>\finTrou
            \node at ({#1+#2/2},-3mm) {$2^{\ik}$};
        \else
            \ifnum \ik<7
                \node at ({#1+#2/2},-3mm) {$\frac{K_{max}}{2^2}$};
            \else \ifnum \ik<8
                \node at ({#1+#2/2},-3mm) {$\frac{K_{max}}{2}$};
            \else
                \node at ({#1+#2/2},-3mm) {$K_{max}$};
            \fi \fi
        \fi
    }
    \ifnum #5>1
        \IPOPdiffKAuxAtrou{#1+#2+#3}{2*#2}{#3}{#4}{\numexpr#5-1\relax}{#6+1}
    \fi
}
\newcommand{\IPOPdiffKAtrou}{
    \pgfmathsetmacro \hauteur{0.5}
    \pgfmathsetmacro \largeur{7}
    \pgfmathsetmacro \ecart{0.4}
    \pgfmathsetmacro \nRectangles{9}

    \pgfmathsetmacro \debutTrou{4}
    \pgfmathsetmacro \finTrou{5}

    \pgfmathsetmacro \largeurDesc{ \largeur / 2^(\nRectangles)}
    \IPOPdiffKAuxAtrou{0}{\largeurDesc}{\ecart}{\hauteur}{\nRectangles}{0}

    \node at (2.3,3mm) {\LARGE $\dots$};
    \node at (2.3,-3mm) {\LARGE $\dots$};
}

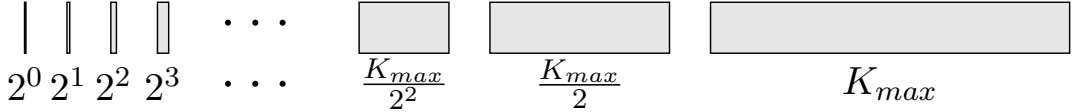
\begin{figure}[t] 
\centering
\scalebox{1.35} {
\begin{tikzpicture}
    \IPOPdiffKAtrou
\end{tikzpicture} }
\caption{Illustration 
of the K-Distributed algorithm.}
\label{fig:kdistr}
\end{figure}

This leads us to consider running concurrently all descents with a distinct $K$ value each. 
As illustrated on Figure \ref{fig:kdistr}, the  $\log_2 (K_{max}) + 1$ descents are executed at the same time, each with a population size equal to $K \times \lambda_{start}$, for $K = 2^0,2^1,2^2, .. ,K_{max}$.
We refer to this algorithm as `K-Distributed`. We implement it with MPI by splitting the initial communicator into $\log_2 K_{max} + 1$ sub-communicators, each containing twice as many processes as the previous one.

\tempnewpage

\section{Experimental results \pierre{on Fugaku}}
\label{section:results}
\bilel{In this section, we conduct a step-by-step performance analysis \pierre{on the supercomputer Fugaku} of the different 
parallel \pierre{and high performance} strategies described before. We first start describing our experimental set-up including the setting of the optimization benchmark functions, the considered parallel computing environment \pierre{and how our parallel strategies are specified for the Fugaku hardware}. Then, we report the benefits of using \pierre{sequential and multi-threaded} BLAS/LAPACK routines 
for CMA-ES. 
Our main results dealing with the different parallel strategies are then reported and analyzed in a comprehensive manner.}

\subsection{Experimental setup}
\label{s:setup}
\bilel{Firstly, for the purpose of comparing 
the considered algorithms from a pure optimization perspective, and for benchmarking purposes,} we consider the COmparing Continuous Optimizers (COCO)~\cite{HansenTMAB16} framework providing an implementation of the Black-Box Optimization Benchmarking (BBOB)\footnote{See also:  \url{https://numbbo.github.io/data-archive/bbob/} } test suite~\cite{Hansen10real-parameterblack-box}.
BBOB is a state-of-the-art blackbox test suite, used in particular in a reference 
workshop held yearly in the well-established Genetic and Evolutionary Computation Conference (GECCO). More than 200 algorithms were already benchmarked within this framework. BBOB provides a set of 24 continuous functions exposing different properties believed to represent a \bilel{relatively} broad range of blackbox optimization problems that one may encounter in practice. These functions are organized into five groups of increasing difficulty \bilel{in terms of separability, multi-modality, illness, conditioning, etc. The first group of functions ($f_1$ to $f_5$) are separable. The second group contains functions of low or moderate conditioning ($f_6$ to $f_9$), while the third group contains unimodal functions with high conditioning ($f_{10}$ to $f_{14}$). The fourth and fifth group contain multi-modal functions with respectively adequate ($f_{15}$ to $f_{19}$) and weak ($f_{20}$ to $f_{24}$) global structure. All functions are available in multiple dimensions: \pierre{in this paper we will study dimensions 10, 40, 200 and 1000}. It is to notice that algorithms benchmarked using the BBOB functions, and the underlying COCO platform, usually fix as a budget the total amount of function evaluations that an algorithm is allowed to query. This is in fact a common practice in a blackbox optimization scenario, where the number of function evaluations is considered to be critical. In other words, this allows one to fairly evaluate the ability of an algorithm to search for a high-quality solution, when facing a range of optimization problems with different structural properties and dimensions, while using the minimum number of blackbox function evaluations. As such, although the time it takes to query one blackbox function evaluation may vary across different BBOB functions of different dimensions, the BBOB test suite does not allow to explicitly control the evaluation times, \pierre{ which are in fact very short  
(less than \david{9ms in dimension 1000} on average across all functions).}
In practice however, function evaluation time directly impacts the overall CPU time required to run an algorithm using a specified budget in terms of the total number of function evaluations. Importantly, the time to evaluate a blackbox function is an important feature to account for when fairly assessing the performance of a parallel optimization algorithm, since it can directly impact 
\pierre{the computation grain size (i.e. the amount of work performed by each "task" in parallel).}
Therefore, in our work, and in addition to the broad range of functions provided by the BBOB test suite, we also consider to accommodate different 
blackbox evaluation times \pierre{by adding \david{artificial} 
additional times to the BBOB function evaluations}. 
\david{For dimensions 10 and 40, for which BBOB functions have low evaluations costs, we will hence also study additional costs of 1ms, 10ms and 100ms.}
This will allow us for a more comprehensive and realistic performance assessment of the considered parallel algorithms.
}
Note that having such greater evaluation times, 
even for 
small dimensions, is easily encountered in function optimization. 
For instance, Roussel et al.\cite{Roussel_Lemaire},
use CMA-ES for \david{parameter estimation} with 
objective functions of dimension 8 and evaluation costs in the order of magnitude of the second. Evaluation costs may be even larger when scientific simulations or neural networks are involved:
for example, using CMA-ES to find hyper-parameters for neural network training can necessitate evaluation times of 
5 or 30 minutes in dimension 19\cite{CMA_neural}.
\david{ Other examples of 
evaluation times include:
groundwater bioremediation (about 8 minutes)\cite{groundwater},
aerodynamics of a shape (about 3 or 11 minutes)\cite{CFD_aerodynamics},
molecular docking (4 hours)\cite{molecular_docking},
automotive crash simulation (about 17 or 29 hours)\cite{CSM_crash} ;
and neural network trainings for 
computer vision (0 to 30 minutes)\cite{hyperparam_vision} or for 
document classification (average of 2.5 or 5.8 hours)\cite{multitask_bayes} . }

\bilel{Secondly, for the purpose of studying the parallel performance of our algorithms 
\pierre{on large‐scale parallel architectures with thousands of CPU cores,} we consider running our 
algorithms on top of the supercomputer Fugaku.}  
\pierre{In June 2024, Fugaku was} the fourth most powerful supercomputer in the 
TOP500 list\cite{top500}, and the first in the world in the 
HPCG list\cite{hpcg} and in the GRAPH500 BFS list\cite{graph500}.
This massively parallel supercomputer contains 158,976 A64FX CPUs,  
which are 
ARM-based architectures developed by Fujitsu with 48 compute cores and 4 assistant cores each   \cite{A64FX,FugakuA64FX}.
The A64FX is divided in 4 CMGs (core memory groups) of 12 \pierre{cores each.} 
Each CMG is a NUMA (non-uniform memory access) node.
Indeed, the memory space of the CPU consists of a set of HBM2 high-bandwidth memory and 2 levels of cache for each of the 4 CMGs. The CMGs communicate between each others and with the network using a ring bus.
The A64FX CPUs are connected by the Tofu Interconnect D \cite{FugakuTofu}, a 6D torus topology.
For our experiments, we consider using 128 A64FX CPUs of the Fugaku, hence representing a total of 512 CMGs and 6,144 compute cores. 

\david{Regarding our implementations, they are all 
based on the sequential C reference code of CMA-ES\footnote{See: \url{https://github.com/cma-es/c-cmaes}}, which we modified for the BLAS/LAPACK routines and the MPI+OpenMP parallelization. 
They are compiled with the Fujitsu C Compiler (version 4.11.1), the Fujitsu OpenMP and MPI libraries, 
and the Fujitsu thread-parallel implementation of BLAS/LAPACK. 
%
We let the C reference code 
set default values for all parameters, except for the initial mean $m$, the initial variance $\sigma$ and $\lambda_{start}$. In order to better adapt to the BBOB search space, we indeed set at the start of each CMA-ES descent the initial mean $m$ to a point selected uniformly at random in the BBOB search space, 
and the initial variance $\sigma$ to 1/4 of the search space width.
}
\pierre{Regarding $\lambda_{start}$, the usual setting is of the order of magnitude of ten.
In order to obtain the best parallel speedups and to compare our strategies on a large number of CPU cores within a single 
setting, we target the processing of each evaluation on a dedicated CPU core (see Section \ref{s:parallel_descent}). For our MPI+OpenMP performance tests on Fugaku, we thus choose to have $\lambda_{start}=12$. This way we can have $T=12$ threads in each MPI process: the $K \times \lambda_{start}$ evaluations of a $K$ descent are thus performed with $K$ MPI processes with $T$ threads each.
Each A64FX runs 4 such MPI processes, i.e. one per CMG as usually done with NUMA architectures.  
For K-Replicated we set $K_{max}$ to  $2^9$ which leads to $K_{max} \times \lambda_{start} = 2^9 \times 12 = 6144$ parallel evaluations executed on 6,144 cores (512 CMGs, 128 A64FX) for the final descent.
For K-Distributed, we set $K_{max}$ to $2^8$ which leads to $(\sum_{i=0}^{8} 2^i) \times \lambda_{start} = 511 \times 12 = 6132$ parallel evaluations on 511 CMGs. The K-Distributed strategy thus uses 12 fewer cores than the K-Distributed one, but this is the fairest comparison we can make between these two strategies.}
%
\bilel{We let the sequential IPOP-CMA-ES execute with $K_{max}=2^9$, and we ensure that this is the sole process running on the CMG of its core, so as to prevent cache interference with other processes.
To keep our experiments manageable in a reasonable time, }
\david{ the execution limit of all experimented algorithms is 12 hours of wall-clock 
time. }
%
\david{ 
Except for Section~\ref{sectionLinalgResults} (BLAS/LAPACK tuning), we conducted 20 runs for each function and each strategy 
in dimensions 10 and 40. Due to time constraints, we performed at least 5 runs for each in dimensions 200 and 1000. The results presented in these sections are aggregated over these multiple runs, with the aggregation methods detailed later. }

\tempnewpage

\subsection{Linear algebra performance results}
\label{sectionLinalgResults}

\begin{figure}[t!] 
\includegraphics[page=1,width=0.38\linewidth]{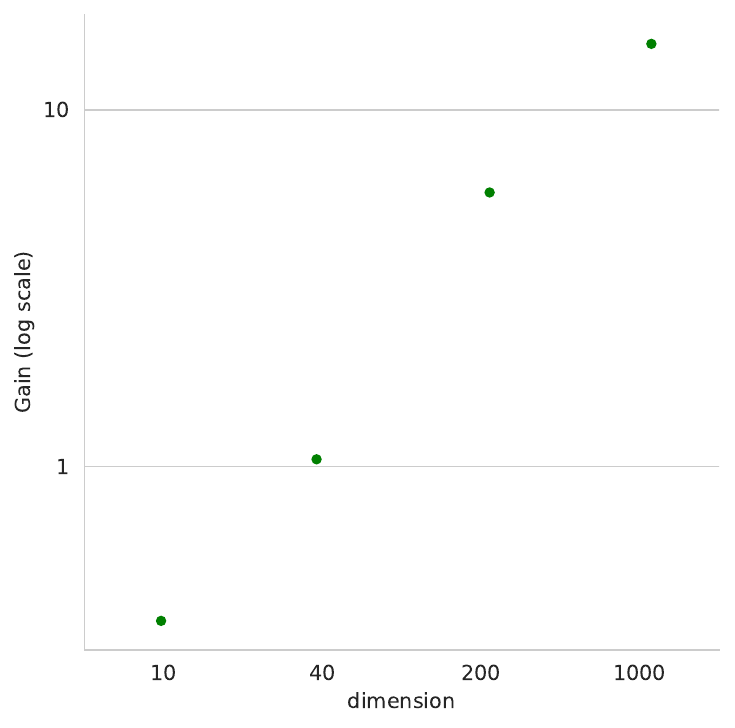}
\if\HAL1
    \hspace{0.8cm}
\else
    \hspace{1.2cm}
\fi
\includegraphics[page=2,width=0.45\linewidth]{images/figures_BLAS_Lapack.pdf}\\
\includegraphics[page=3,width=0.45\linewidth]{images/figures_BLAS_Lapack.pdf}
\includegraphics[page=4,width=0.5\linewidth]{images/figures_BLAS_Lapack.pdf}
\caption{(upper-left) Performance gains \pierre{for the eigendecomposition of the $C$ matrix when using LAPACK over the reference C code (written without LAPACK)}. 
(upper-right, resp. lower-left) \pierre{Performance gains for the adaptation of the $C$ matrix (resp. for the sampling) when using Level 2 or Level 3 BLAS over the reference C code (without BLAS).}  
(lower-right) \pierre{Performance gains over the reference C code (without BLAS and LAPACK) for all the linear algebra part, with LAPACK for the eigendecomposition and Level 3 BLAS for the $C$ matrix 
adaptation, when using Level 2 or Level 3 BLAS routines  
for the sampling.}
\david{ 
 The IPOP columns correspond to a IPOP-CMA-ES execution with successive descents using 
 $K$ 
 from 1 to $2^8$.
 }
}
\label{gainsBLASLAPACK}
\end{figure}

\pierre{We start our analysis by studying the performance impact of introducing BLAS/LAPACK routines 
in three linear algebra steps as described in Section \ref{s:blas_lapack}. 
Figure~\ref{gainsBLASLAPACK} presents the performance gains with respect to each step,  
each step being executed sequentially for various dimensions and for various $K$ values, 
with $\lambda_{start} = 12$. 
}

\bilel{More precisely}, the upper-left sub-figure reports the performance gain of using LAPACK specifically with respect to the eigendecomposition operation in CMA-ES. 
LAPACK enables 
us to accelerate the eigendecomposition step for problems of dimensions 40 and above, and significantly for problems of dimensions 200 et 1000 (up to 15.3x), where the 
$C$ matrix is large enough 
to benefit from the LAPACK performance optimisations. 
\bilel{Notice that for a relatively small dimension of 10, we found that using LAPACK leads to a performance loss, which is 
because the matrices maintained by CMA-ES are so small for such a dimension}. However, since in dimension 10 the eigendecomposition accounts for only 9\% of the overall linear algebra runtime (averaged over the 24 functions of BBOB), \bilel{the overall loss in performance} is negligible. 
\pierre{The upper-right part of Figure~\ref{gainsBLASLAPACK} presents BLAS performance gains with respect to the  adaptation of the $C$ covariance matrix. 
We distinguish here gains obtained when directly using Level 2 BLAS in equation \ref{eq:2} (see Section \ref{s:pres_cmaes}), and gains obtained with Level 3 BLAS thanks to our rewriting proposed in Section \ref{s:blas_lapack}.}   
\pierre{While the use of Level 2 BLAS does not offer performance gains in dimensions 10, 40 and 200, 
our new computation scheme based on Level 3 BLAS offers very significant performance gains (up to 190x), especially for higher problem dimensions. The extra affectations (see Section \ref{s:blas_lapack}) are thus offset by the BLAS gain, even for the lowest dimension. }
\bilel{As for the sampling operations, as reported in the lower-left sub-figure, we observe that using Level 2 routines \pierre{directly in equation \ref{eq:1} (see Section \ref{s:pres_cmaes})} can only provide some gain for dimensions greater than 10.}
However, when the operations are rewritten using Level 3 routines (see Section \ref{s:blas_lapack}), we are able to \pierre{accelerate the reference C code for any dimension, with gains stronger than for Level 2 BLAS. Again the extra affectations (see Section \ref{s:blas_lapack}) are offset by the BLAS gains. }
\david{
Finally, in the lower-right part of Figure~\ref{gainsBLASLAPACK}, we report performance gains due to the sampling operations, but this time in a different context. The gain is computed with respect to \emph{all} the linear algebra 
\pierre{part (i.e. both sampling at lines \ref{a:sampling_start}-\ref{a:sampling_end} and update at line \ref{a:update} in Algorithm \ref{algo:cmaes}), 
and not just to 
the sampling step 
like in the lower-left sub-figure.}
The eigendecomposition uses LAPACK and the matrix adaptation uses Level 3 BLAS, \pierre{whereas Level 2 or Level 3 BLAS are used for the sampling}. 
}
Although the gains obtained solely for the sampling may be deemed relatively small compared to the ones obtained solely for the covariance matrix adaptation, \bilel{using Level 3 BLAS for the sampling operations still has a relatively substantial impact when LAPACK and BLAS routines are \pierre{already} 
used to optimize the other linear algebra steps}.
For instance, this 
enables us to increase the overall gain \pierre{over the C reference code} from $1.5$ to $2.5$ for all the linear algebra operations in dimension 1000. 
 \pierre{As a final remark, regarding the sampling and the adaptation steps, one can see stronger gains for $K=2^8$ and for IPOP-CMA-ES than for $K=1$: this is due to the larger population sizes, which lead to larger matrices.
 This shows that BLAS/LAPACK routines are even more relevant for IPOP-CMA-ES than for CMA-ES, the IPOP-CMA-ES increasing population sizes becoming eventually larger than the CMA-ES ones.} 


\begin{table}[t!] 
\centering
\caption{Proportions \pierre{(averaged over all BBOB functions)} of the linear algebra runtime within the overall runtime of a \pierre{sequential} execution of IPOP-CMA-ES (with $\lambda_{start} = 12$ and $K_{max}=2^8$)
.}
\label{propoLA}
\begin{tabular}{c|cccc}
& \multicolumn{4}{c}{Dimension} \\
\bilel{Level 3 BLAS / LAPACK} & 10 & 40 & 200 & 1000 \\
no & 38\% & 36\% & 44\% & 69\% \\
yes & 33\% & 21\% & 18\% & 21\% \\
\end{tabular}
\end{table}

\bilel{To further illustrate the impact of linear algebra operations,
Table~\ref{propoLA} presents the proportion of CPU time these operations consume relative to the total execution time of IPOP-CMA-ES, with and without 
Level 3 BLAS / LAPACK routines.}
\david{ As we can see, BLAS/LAPACK becomes increasingly effective at reducing the proportion of linear algebra computations as the dimension increases. High dimensionnality is a key factor that makes an optimization problem hard, making our linear algebra rewrites particularly valuable for tackling harder problems. }
\pierre{Thanks to our BLAS/LAPACK rewrites, the linear algebra part is now minority in the overall IPOP-CMA-ES runtime. Using additional costs for the evaluations (see Section \ref{s:setup}) will make the linear algebra part even more minority.}

\pierre{Finally, }\bilel{we tuned our BLAS/LAPACK implementations to determine the optimal number of threads 
(up to a maximum of 12) for each dimension}. We found that dimensions 10 and 40 run best with 1 thread, dimension 200 with 4 threads and dimension 1000 with 12 threads. \bilel{This aligns with the observation that larger dimensions entail larger matrices, which can be effectively managed by BLAS/LAPACK with a greater number of threads.}
\pierre{However, the sizes of the involved matrices are not large enough, and} \david{the best speedup we obtain for running BLAS/LAPACK on multiple threads is $1.4\times$, for dimension 1000 with 12 threads.} \pierre{Due to this limited speedup , we believe that we would not benefit from distributing the linear algebra operations over multiple MPI processes (i.e. over more than 12 cores). 
That is why we chose to perform the linear algebra operations in parallel using only multi-threading within one MPI process (the main one for each descent, see Section \ref{s:parallel_descent}), and up to 12 threads.}

\pierre{Now that we have accelerated the linear algebra operations, with BLAS/LAPACK routines 
and as much as possible via parallelism,} \bilel{the time spent on function evaluations is the majority 
in the IPOP-CMA-ES execution time;   
hence, making the relevant 
parallelization of function evaluations of high importance. \pierre{We will thus now} 
focus on our two proposed parallel strategies.}

\tempnewpage

\subsection{Parallel performance results}

\bilel{In this section, we delve into the performance behavior of the proposed K-Replicated and K-Distributed parallel algorithms. A thorough and fair performance assessment of our parallel variants requires to discuss two aspects. Firstly, although the function evaluation time can significantly influences the overall performance, it cannot be explicitly controlled in the COCO implementation of the BBOB functions. Hence, one has to adopt a more robust approach to assessing parallel performance relative to function evaluation time. Secondly, due to the stochastic nature of the considered algorithms, K-Replicated and K-Distributed are not expected to deliver exactly the same output as the sequential IPOP-CMA-ES. Consequently, it is essential to carefully define a metric that allows us to fairly evaluate the ability of the different algorithms to reach high-quality solutions within reduced time-frames. Therefore, in the next subsection, we begin by discussing the methodology we adopt to address these two aspects. Subsequently, we present our findings and state our main results.}
\subsubsection{Performance assessment methodology}


\if\HAL1
    \begin{figure}
    \centering
    \includegraphics[width=0.6\textwidth]{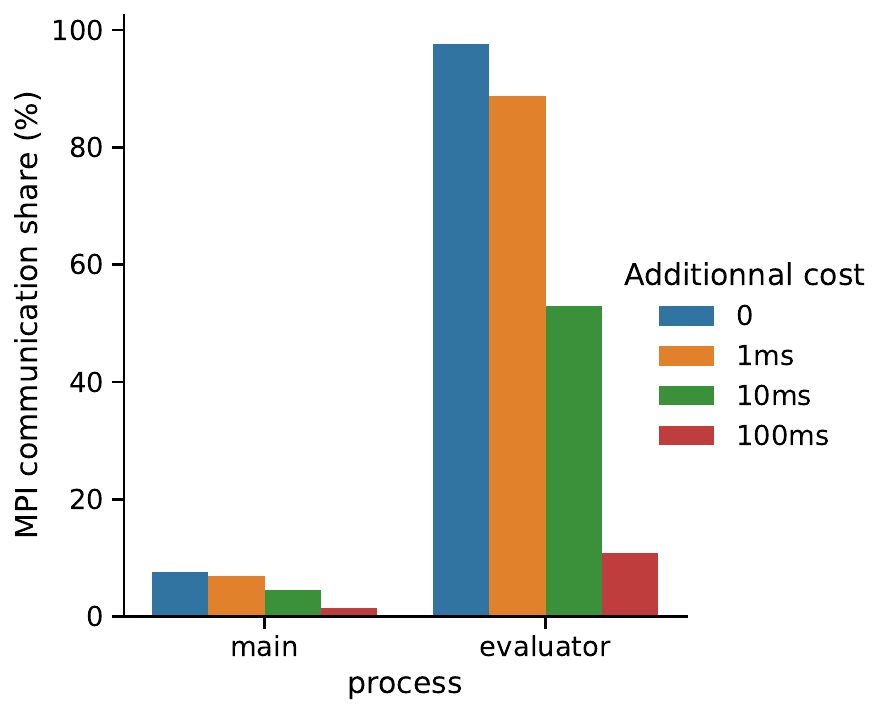}
    \caption{
    MPI communication shares with respect to the total runtime, as measured by the Fugaku Instant Performance Profiler (FIPP) \cite{fipp} for a $K=2^8$ descent with 256 MPI processes, 
    averaged over all BBOB functions of dimension 40.
    `main` is the main MPI process (namely, the one with rank 0) driving the $K=2^8$ descent (see Section \ref{s:parallel_descent})
    and processing the linear algebra operations (see end of Section \ref{sectionLinalgResults}), 
    whereas `evaluator` is one of the MPI processes performing only evaluations (here, the one with the highest MPI rank). Contrary to Table \ref{propoLA}, we consider only $K=2^8$ and all the evaluations are performed in parallel. 
    }
    \label{fig:FIPP}
    
    \centering
    \includegraphics[page=1,width=0.8\textwidth]{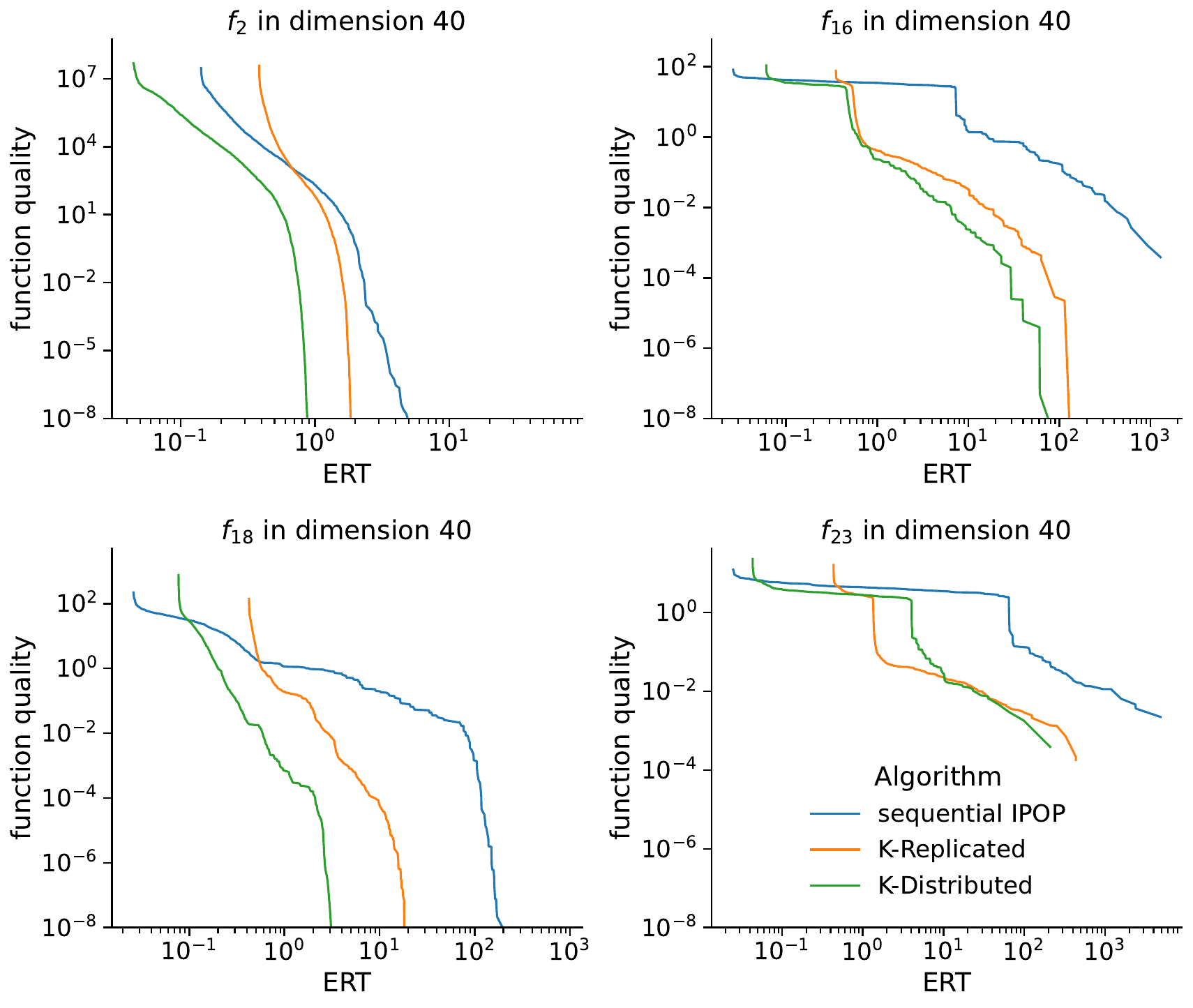}
    \caption{ \david{ Function quality of the best solution found over runtime,
    averaged over 20 runs using the Expected Runtime (ERT) \cite{HansenERT}.
    } }
    \label{fig:convergExample}
    \end{figure}
\else
    \begin{figure}
    \centering
    \begin{minipage}[b]{.4\textwidth}
        \centering
            \includegraphics[width=1\linewidth]{images/figures_fipp.pdf}
        \caption{
        MPI communication shares with respect to the total runtime, as measured by the Fugaku Instant Performance Profiler (FIPP) \cite{fipp} for a $K=2^8$ descent with 256 MPI processes, 
        averaged over all BBOB functions of dimension 40.
        `main` is the main MPI process (namely, the one with rank 0) driving the $K=2^8$ descent (see Section \ref{s:parallel_descent})
        and processing the linear algebra operations (see end of Section \ref{sectionLinalgResults}), 
        whereas `evaluator` is one of the MPI processes performing only evaluations (here, the one with the highest MPI rank). Contrary to Table \ref{propoLA}, we consider only $K=2^8$ and all the evaluations are performed in parallel. 
        }
        \label{fig:FIPP}
    \end{minipage}
    \hfill 
    \begin{minipage}[b]{.55\textwidth}
        \centering
        \includegraphics[page=1,width=1\linewidth,trim=0 0 0 0cm,clip]{images/conv_ex_seq_krep_kdis.pdf}
        \caption{ \david{ Function quality of the best solution found over runtime,
        averaged over 20 runs using the Expected Runtime (ERT) \cite{HansenERT}.
        } }
        \label{fig:convergExample}
    \end{minipage}
    \end{figure}
\fi

\textit{\textbf{Function evaluation time.}}
\pierre{We first emphasize the relevance of the additional costs introduced in Section \ref{s:setup}, 
by reporting 
in Figure~\ref{fig:FIPP} the share of MPI communications in the total runtime of a $K=2^8$ descent involving 256 MPI processes. 
When considering zero additional cost, the time spent in MPI communications (scatter and gather operations, see Section \ref{s:parallel_descent}) is limited for the main process, but is the vast majority of the total time for an  
other process involved in the parallel descent. 
This is due to the linear algebra part, only performed in the main process and leading to important waiting times for processes other than the main one in their MPI communications (namely at the scatter level).
The linear algebra part can thus be a potential performance bottleneck when scaling on a large number of CPU cores. }

\pierre{When adding extra costs in the function evaluation, one can see in Figure~\ref{fig:FIPP} that the MPI communication shares (i.e. the relative cost of the linear algebra part with respect to the total time, as well as the data transfers themselves) strongly decrease with an increasing additional cost, until becoming a minority. These extra costs enable thus us to also simulate real-life cases (see Section \ref{s:setup}) where the linear algebra is not 
a performance bottleneck. 
}


\tempnewpage


\noindent \textit{\textbf{\bilel{Solution quality.}}} \bilel{Since the optimal solutions of the BBOB functions are known, we can evaluate the quality of a solution relative to the optimal one. In our work, we measure quality by the difference $\epsilon$ between the function value of the best solution found so far by an algorithm and the function's optimal value. However, since the considered algorithms are stochastic, we get inspiration from the so-called Expected Runtime (ERT)~\cite{HansenERT} in order to aggregate the quality results obtained from multiple runs using different seeds for a same algorithm. More specifically, the ERT defines the empirical average time (over the different seeds) it takes for an algorithm to hit a solution of a given target quality $\epsilon$. Notice that two scenarios can happen. In the case all the runs of an algorithm were successful to hit the target quality $\epsilon$, the ERT is simply the average (over all runs) of the hitting times. In the case some runs were unsuccessful in hitting a target quality within the maximum affordable budget, we can assume that the algorithm could have been restarted for a new run until a success is observed (which is typically the case for stochastic algorithms such as ours). The time until a new run is successful can then be viewed as a random variable, which average value can be empirically estimated in a straightforward manner using the data available from the (other) successful runs at hand. Hence, the ERT value is simply defined as the sum of the execution times of all runs (including unsuccessful ones) divided by the number of successful runs. Notice that for the ERT to be correctly defined, at least one run must be successful. The reader is referred to \cite{HansenERT} for more details.}

\bilel{In all our algorithms, it is easy to track the quality of the best solution found so far over time. Hence, we can easily compute the ERT 
of an algorithm when considering different targets. Consequently, we can report the expected convergence profile of an algorithm, as shown in Figure~\ref{fig:convergExample} on four illustrative BBOB functions. The reported convergence profile suggest a number of important issues to be carefully considered when assessing the relative performance of algorithms. Firstly, we can see that the relative behavior of the three algorithms depends on the tackled function, i.e., no algorithm is better than all others for all functions. Importantly, the relative performance of an algorithm depends on the considered target quality. We can also see that \emph{not} all algorithms can hit the same range of targets on \emph{all} functions. Some algorithms are even not able to reach the same target that other algorithms can reach for the same function. Consequently, such observations raise a number of qualitative and quantitative questions, i.e., which algorithm is able to reach a given quality? which algorithm reaches it faster than the others ? what relative speedup can be obtained for a given solution quality? etc. In particular, parallel speedups cannot be defined in the conventional manner used in parallel computing. Instead, in our work, we consider nine fixed target quality values, namely $\epsilon\in \{10^{2}, 10^{1.5}, 10^{1},10^{0.5},10^{0},10^{-2},10^{-4},10^{-6},10^{-8}\}$. These values are actually the same as the ones used in the COCO framework~\cite{HansenTMAB16}. Roughly speaking, this is intended to represent a range of target quality going from easy, to moderate and difficult to achieve. For each given pair of BBOB function and target quality $\epsilon$, we can then fairly analyze the relative ERTs achieved by the three algorithms. More specifically, in the following section, we define the speedup achieved by an algorithm over another one with respect to a given BBOB function and a given target quality $\epsilon$, as the ratio of the ERTs achieved by the two compared algorithms.}

\tempnewpage

\subsubsection{Overall parallel speedup}
\label{subsec:overall-speedups}


\newcommand{\tableMultiDimCost}{
\begin{tabular}{c|c|c|c|c|c|c|c|c|c|c}
Dimension & 10 & 10 & 10 & 10 & 40 & 40 & 40 & 40 & 200 & 1000 \\
Additional cost & 0 & 1ms & 10ms & 100ms & 0 & 1ms & 10ms & 100ms & 0 & 0 \\
\hline
{\bf K-Replicated} &&&&&&&&&& \\
avg. speedup & 1.1 & 83 & 159 & 219 & 8.6 & 70 & 160 & 176 & 59 & 23 \\
std. dev. & 4.0 & 234 & 432 & 639 & 24 & 171 & 520 & 618 & 183 & 55 \\
min. speedup & 0.1 & 0.1 & 0.6 & 3.7 & 0.0 & 0.1 & 0.5 & 3.9 & 0.1 & 0.1 \\
max. speedup & 30 & 1620 & 2995 & 5018 & 206 & 1182 & 3861 & 5121 & 1614 & 425 \\
\hline
$<$/$>$ & 13/182 & 15/182 & 10/188 & 23/170 & 9/173 & 9/174 & 13/170 & 35/148 & 28/134 & 20/107 \\
\hline
{\bf K-Distributed} &&&&&&&&&& \\
avg. speedup & 2.7 & 115 & 201 & 169 & 17 & 171 & 419 & 736 & 392 & 70 \\
std. dev. & 3.2 & 259 & 378 & 255 & 39 & 302 & 799 & 2884 & 1580 & 257 \\
min. speedup & 0.5 & 0.6 & 2.1 & 7.7 & 0.3 & 0.7 & 1.8 & 8.0 & 0.3 & 0.5 \\
max. speedup & 20 & 1610 & 1901 & 2071 & 267 & 1645 & 3857 & 18080 & 13944 & 2397 \\
\end{tabular}}

\begin{table}[t] 
\centering
\caption{
\david{ Speedups obtained by the two parallel strategies over the sequential IPOP‐CMA‐ES, aggregated over the different pairs of BBOB functions and target qualities, for various function dimensions and function granularities.
'<' (resp. '>') stands for the number of function-target couples for which K-Replicated reaches the target quality before (resp. after) K-Distributed. \pierre{Function-target} couples are accounted for when both parallel algorithms reach the target quality, so the sum of the counts may vary with the dimension and the granularity. }
}
\if\HAL1
    \scalebox{0.85}{\tableMultiDimCost}
\else
    \tableMultiDimCost
\fi
\label{table:multiDimCost}
\end{table}

\bilel{In Table~\ref{table:multiDimCost}, we summarize some basic statistics concerning the speedups obtained by our two parallel strategies over the sequential IPOP-CMAE-ES. \pierre{This sequential IPOP-CMA-ES leverages our Level 3 BLAS / LAPACK rewrites (with one thread) so as to focus here on the speedups obtained thanks to MPI+thread parallelism.} 
More precisely, Table~\ref{table:multiDimCost} reports the average, the standard deviation, the minimum and the maximum, speedups obtained respectively by K-Replicated and K-Distributed over the different pairs of BBOB functions and target qualities. The results are reported for each considered function dimension and function granularity.
\david{ 
Additionally, the row "</>" of Table~\ref{table:multiDimCost} offers a direct comparison : each function-target pair is counted in the left-hand (resp. right-hand) number when K-Replicated reaches the target quality before (resp. after) K-Distributed.}
For example, the first cell with value 13/182 reads as K-replicated has better ERT than K-Distributed on 13 function-target pairs, whereas K-Distributed has better ERT than K-replicated on 182 function-target pairs.
Two main observations can be extracted from Table~\ref{table:multiDimCost}.}

\bilel{Firstly, we can clearly see that K-Distributed \pierre{(despite using a little less CPU cores)} provides almost always a better average speedup than K-Replicated. The only exception is for dimension 10 with 100ms additional cost. Interestingly, even in such a case, as shown in the "</>" row, the number of function-target pairs where K-Distributed is faster than K-Replicated is substantial. In fact, this holds true with no exception independently of the setting of the dimension and of the function additional cost. 
\pierre{Moreover, we managed to obtain very high maximum speedups 
(over functions and targets) 
for both strategies, as soon as the computation grain or the dimension are large enough.  
It is also interesting to note that we even obtain 'super-linear' speedups for K-Distributed.} This happens for dimension 40 with a function additional cost of 100ms, as well as for dimension 200, where the maximum observed speedup of K-Distributed is respectively \david{18080$\times$ for function $f_7$ with target $10^{-6}$, and 13944$\times$ for $f_{18}$ with $10^{-2}$}, which is substantially larger than the 6144 cores used. 
One should indeed recall that the considered parallel strategies do \emph{not} imply exactly the same search behavior as  serial IPOP-CMAE-ES. Hence, these results support the fact that, for some specific functions/targets, the considered parallel strategies are able to show better search behavior than the original serial algorithm.}
\bilel{
To further illustrate the superiority of K-Distributed, we show in Table~\ref{table:KdisKrep} a detailed view of the corresponding speedups obtained over K-Replicated for dimension 40 and an additional evaluation cost of 100ms. We clearly observe that K-Distributed is the faster solver for most targets. Notice that the \pierre{relative} speedup on function 7 is 
very high, \pierre{K-Distributed being more than $500\times$ faster than K-Replicated,} which is because 
K-Replicated is particularly inefficient for this function (as further detailed in Section \ref{section:diffk}). Besides, it is interesting to note that not all targets can be hit by the algorithms for any function which is specifically because some BBOB functions are more difficult to optimize than others. For the clarity of the presentation, a more thorough discussion of this important aspect is delayed to later.} 




\newcommand{\tableKDistrVSkReplPres}{\resizebox{0.5\columnwidth}{!}{
\begin{tabular}{ c | c|c|c|c|c|c|c|c|c }
& \multicolumn{9}{c}{targets} \\
function & \multicolumn{1}{c}{$10^{2}$} & \multicolumn{1}{c}{$10^{1.5}$} & \multicolumn{1}{c}{$10^{1}$} & \multicolumn{1}{c}{$10^{0.5}$} & \multicolumn{1}{c}{$10^{0}$} & \multicolumn{1}{c}{$10^{-2}$} & \multicolumn{1}{c}{$10^{-4}$} & \multicolumn{1}{c}{$10^{-6}$} & \multicolumn{1}{c}{$10^{-8}$} \\
 \hline
1 & 0.6 & 0.9 & 0.9 & 1.0 & \bf1.0 & \bf1.2 & \bf1.3 & \bf1.4 & \bf1.4 \\
2 & \bf7.5 & \bf9.0 & \bf10 & \bf11 & \bf12 & \bf14 & \bf14 & \bf13 & \bf13 \\
3 & \bf1.2 & \bf5.7 & X & - & - & - & - & - & - \\
4 & \bf1.4 & \bf7.7 & - & - & - & - & - & - & - \\
5 & \bf1.4 & \bf1.6 & \bf1.8 & \bf1.9 & \bf2.0 & \bf2.2 & \bf2.3 & \bf2.3 & \bf2.4 \\
6 & \bf1.0 & \bf1.1 & \bf1.3 & \bf1.4 & \bf1.5 & \bf1.9 & \bf2.2 & \bf2.4 & \bf2.4 \\
7 & 1.0 & \bf1.1 & \bf1.3 & \bf49 & \bf495 & \bf508 & \bf508 & \bf508 & \bf495 \\
8 & \bf1.1 & 0.9 & \bf2.6 & \bf2.7 & \bf2.7 & \bf2.9 & \bf3.0 & \bf2.9 & \bf2.9 \\
9 & \bf1.2 & 0.8 & \bf3.2 & \bf3.4 & \bf3.5 & \bf3.7 & \bf3.9 & \bf3.9 & \bf3.8 \\
10 & \bf8.7 & \bf9.5 & \bf11 & \bf12 & \bf13 & \bf15 & \bf15 & \bf15 & \bf14 \\
11 & \bf24 & \bf22 & \bf22 & \bf22 & \bf22 & \bf20 & \bf19 & \bf17 & \bf16 \\
12 & \bf1.3 & \bf1.3 & \bf1.2 & \bf1.2 & \bf1.1 & 0.9 & \bf1.4 & \bf2.0 & \bf2.2 \\
13 & \bf1.1 & \bf1.1 & \bf1.2 & \bf1.2 & \bf1.2 & \bf1.5 & \bf2.5 & \bf5.7 & \bf7.2 \\
14 & \bf2.0 & 0.6 & 0.8 & 0.9 & 1.0 & \bf1.3 & \bf4.2 & \bf9.7 & \bf16 \\
15 & \bf1.2 & \bf5.9 & \bf11 & \bf14 & \bf14 & \bf12 & \bf11 & \bf11 & \bf10 \\
16 & \bf1.7 & \bf1.1 & 1.0 & \bf1.1 & \bf1.5 & \bf50 & \bf23 & \bf26 & \bf29 \\
17 & \bf2.1 & \bf2.2 & 1.0 & 1.0 & \bf1.2 & \bf2.6 & \bf11 & \bf13 & \bf13 \\
18 & \bf1.7 & 0.8 & \bf1.1 & \bf1.3 & \bf1.5 & \bf12 & \bf32 & \bf40 & \bf38 \\
19 & \bf1.5 & \bf2.2 & 1.0 & \bf2.6 & \bf3.1 & - & - & - & - \\
20 & 0.9 & 0.9 & 0.9 & 0.8 & \bf8.1 & - & - & - & - \\
21 & \bf1.4 & 0.8 & 0.8 & 0.7 & 0.1 & 0.1 & 0.2 & 0.2 & 0.2 \\
22 & \bf1.8 & 0.7 & 0.7 & 0.8 & 0.0 & - & - & - & - \\
23 & \bf2.0 & \bf2.0 & \bf1.9 & 0.9 & 0.4 & \bf1.2 & - & - & - \\
24 & \bf1.0 & \bf2.1 & \bf3.7 & \bf4.5 & \bf13 & - & - & - & - \\
\end{tabular}
}}

\begin{table}[t!]
\centering
\caption{ Speedups of K-Distributed over K-Replicated for all targets and functions in dimension 40 with an additional 
cost of 100ms. 
Cells where K-Distributed is faster than K-Replicated (speedup $\geq$ 1) 
are in bold font. 
'X' indicates 
that K-Distributed did not reach the target. '-' indicates 
that no parallel strategy reached the target. 
}
\tableKDistrVSkReplPres
\label{table:KdisKrep}
\end{table}

\bilel{Secondly, we can clearly see \pierre{in Table~\ref{table:multiDimCost}} that the function evaluation granularity as captured by the considered additional costs has a deep impact on the obtained speedups. 
\pierre{Except for one case (when increasing the additional cost from 10ms to 100ms in dimension 10 for K-Distributed),} the speedup of the parallel strategies over the serial algorithm increases indeed consistently with the function evaluation cost. 
}

\bilel{At this stage of the presentation, let us remark that although Table~\ref{table:multiDimCost} provides a global view of the speedups the parallel strategies can achieve over the serial algorithm, the reported statistics are still to be very carefully interpreted.
\david{ In particular, computing a speedup value can only be performed when \emph{both} the sequential IPOP‐CMAE‐ES and the parallel strategy were successful in hitting a given target. }
This means that the average speedup values as shown in Table~\ref{table:multiDimCost} discard the pairs of function/target where at least one algorithm was not able to hit the considered target\footnote{For instance, one can see from Table~\ref{table:KdisKrep} that not all target were hit.}.
Generally speaking, we observed that the harder a target is, the more likely it is for serial IPOP-CMA-ES to be unsuccessful. This is exactly why the overall average speedup values reported in Table~\ref{table:multiDimCost} decreases when going from dimension 
\pierre{200 to dimension 1000}.
However, such a decrease is \emph{not} to be attributed to a parallel performance loss as the problem dimension decreases. It is instead to be attributed to the fact that many target values cannot be hit by serial IPOP-CMA-ES. Hence, a more fine grained assessment of the behavior of the different parallel strategies is needed to fully appreciate the benefits of the designed strategies, in particular as a function of problem dimension. This is to be studied in more detail in the next section introducing more advanced statistics.}

\tempnewpage

\subsubsection{Empirical cumulative distribution analysis}
\label{subsec:ECDF}

\bilel{In this section, we analyze the relative performance of the considered algorithms using the so-called Empirical Cumulative Distribution Functions (ECDF)~\cite{ECDF_data_profils} as introduced in the COCO benchmarking framework\footnote{\url{https://numbbo.github.io/coco-doc/perf-assessment/\#empirical-distribution-functions}}. Generally speaking, an empirical (cumulative) distribution function 
$F : \mathbb{R} \rightarrow [0, 1]$  is defined for a given real-valued data set, such that $F(t)$ equals the fraction of elements in the data which are smaller than or equal to $t$. In an optimization setting, the ECDF is to be viewed as a measure of how many 'problems' a stochastic optimization algorithm can solve on average for a given time budget $t$. More specifically, for the purpose of our analysis, we consider the set containing the (function,target,run)-triplets labeled with the timestamp at which the algorithm was able to find a solution hitting a specified target value. The ECDF then counts for every timestamp $t$ the proportion of (function,target,run)-triplets labeled with a timestamp smaller than or equal to $t$. In other words, the ECDF counts the proportion of targets that an optimization algorithm is able to hit as a function of the elapsed time $t$, i.e., the higher the proportion, the more powerful an algorithm.
}



\begin{figure}[t]
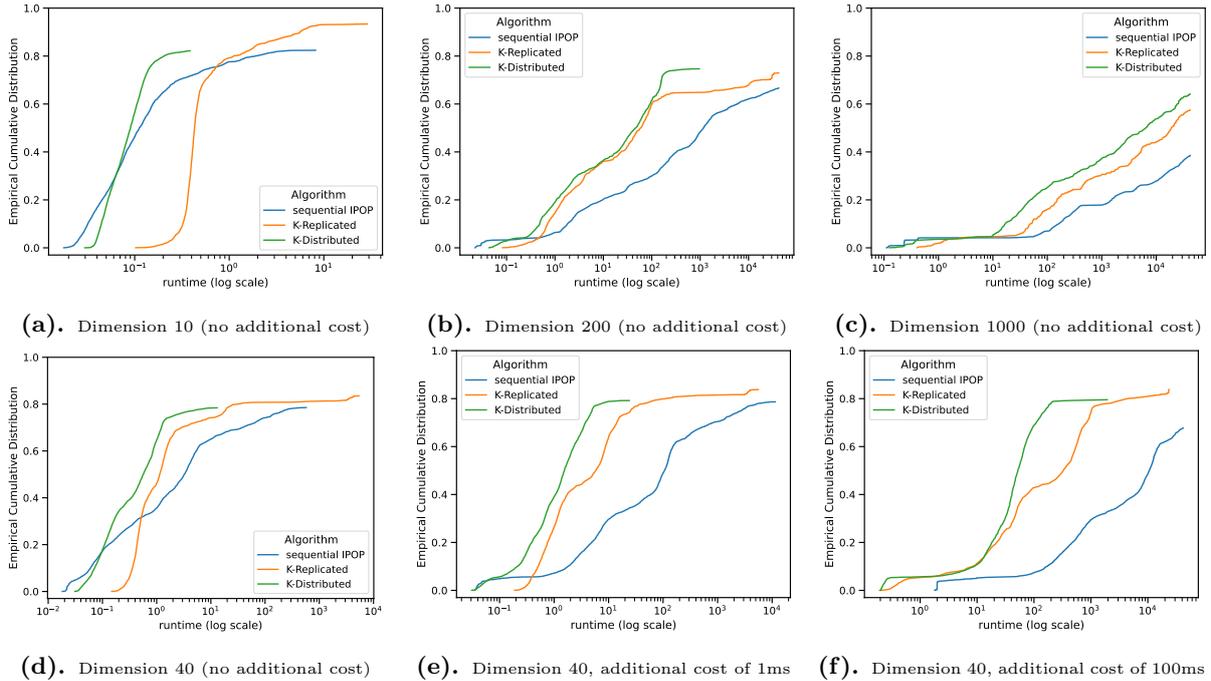


\if\HAL1
    \newcommand{\subfigECDF}[3]{
        \begin{subfigure}[b]{0.32\textwidth}
            \includegraphics[page=#1,width=\linewidth,trim=0 0 0 0cm,clip]{images/figures_A.pdf}
            \caption{\tiny #2}
            \label{fig:#3}
        \end{subfigure}
    }
\else
    \newcommand{\subfigECDF}[3]{
        \begin{subfigure}[b]{0.33\textwidth}
            \includegraphics[page=#1,width=\linewidth,trim=0 0 0 0cm,clip]{images/figures_A.pdf}
            \caption{#2}
            \label{fig:#3}
        \end{subfigure}
    }
\fi

\subfigECDF{9}{Dimension 10 (no additional cost)}{ecdf10}
\subfigECDF{10}{Dimension 200 (no additional cost)}{ecdf200}
\subfigECDF{11}{Dimension 1000 (no additional cost)}{ecdf1000}
\subfigECDF{12}{Dimension 40 (no additional cost)}{ecdf40}
\subfigECDF{13}{Dimension 40, additional cost of 1ms}{ecdf1ms}
\subfigECDF{14}{Dimension 40, additional cost of 100ms}{ecdf100ms}

\caption{ECDF (i.e. rates of function-target couples reached 
for a given runtime) of each algorithm for various dimensions and granularities. 
}
\label{fig:ecdf}
\end{figure}

\newcommand{\tableMultiDimCostReachedWhenKdistrStops}{
\begin{tabular}{c|c|c|c|c|c|c|c|c|c|c}
Dimension & 10 & 10 & 10 & 10 & 40 & 40 & 40 & 40 & 200 & 1000 \\
Additional cost & 0 & 1ms & 10ms & 100ms & 0 & 1ms & 10ms & 100ms & 0 & 0 \\
Sequential IPOP & 72\% & 31\% & 24\% & 21\% & 67\% & 34\% & 34\% & 33\% & 48\% & 39\% \\
K-Replicated & 29\% & 82\% & 83\% & 83\% & 75\% & 74\% & 78\% & 78\% & 65\% & 57\% \\
K-Distributed & 82\% & 82\% & 83\% & 82\% & 78\% & 79\% & 79\% & 80\% & 75\% & 64\% \\
\end{tabular}}

\begin{table}[t] 
\centering
\caption{ECD value reached by each algorithm for the final \pierre{timestamp} 
of K-Distributed for various dimensions and granularities.
}
\tableMultiDimCostReachedWhenKdistrStops
\label{table:reachRate}
\end{table}

\david{
In Figures~\ref{fig:ecdf} \subref{fig:ecdf10} \subref{fig:ecdf200} \subref{fig:ecdf1000} \subref{fig:ecdf40}, we present the ECDF curves for each algorithm across all dimensions (without additional cost).
%
On an ECDF graph, a curve positioned to the left of another one for a given ECD value indicates 
\pierre{a more powerful algorithm}. 
Notably, K-Distributed's curve is almost always the leftmost 
\pierre{which} suggests that, without specific knowledge of a function 
landscape, K-Distributed is the superior choice.
Similarly, 
K-Replicated's curve \pierre{is} to the left \pierre{of the the sequential IPOP-CMA-ES one} for most of the execution time. 
%
\pierre{The ECDF analysis thus} confirms the conclusions from Section~\ref{subsec:overall-speedups}: both parallel strategies outperform the sequential IPOP-CMA-ES, with K-Distributed being the most effective.
}

\david{
Next, we analyze the dynamics of the algorithms with respect to function dimension.
Firstly, higher dimensions show a larger gap between the parallel variants and the sequential IPOP-CMA-ES, \pierre{leading} 
to greater speedups for the parallel variants.
Secondly, for each dimension, there is a cross-over ECD value  
where \pierre{each parallel curve crosses and stays} to the left of the sequential curve, meaning the parallel variant is generally \pierre{better} 
than 
sequential IPOP-CMA-ES for solving problems past this point. \pierre{Sequential IPOP-CMA-ES is therefore faster  only for the very easiest targets and,} as the dimension increases, these 
cross-over values 
decrease, making the parallel variants the better choice for a broader range of problems. 
Thirdly, in Table~\ref{table:reachRate}, we report the ECD values for each algorithm at the final 
\pierre{timestamp} 
of the K-Distributed strategy. We observe that ECD values decrease with increasing dimension, especially at dimensions 200 and 1000. However, for these high dimensions, the parallel strategies show greater ECD values than the sequential IPOP-CMA-ES, \pierre{K-Distributed having the greatest ones}. 
Thus, as dimension increases (which makes optimization problems more challenging), the benefits of our parallel variants, particularly K-Distributed, become more pronounced.
}

\david{
In Figures~\ref{fig:ecdf} \subref{fig:ecdf40} \subref{fig:ecdf1ms} \subref{fig:ecdf100ms}, we report ECDF curves for different granularities at dimension 40
\pierre{(dimension 10 leads to similar results)}. 
%
As with dimension, a higher granularity widens the gap between the parallel strategies and the sequential IPOP-CMA-ES.
Additionally, a higher granularity increases the gap between K-Distributed and K-Replicated for ECD values greater than $0.4$, making K-Distributed a better choice for more \pierre{time-consuming} 
problems.
%
}

\david{
Finally, we observe specific effects in certain dimensions and granularities.
In \pierre{dimension 1000 
(see Figure~\ref{fig:ecdf1000}), the sequential IPOP-CMA-ES stops at a lower ECD value due to the time limit, which prevents us from computing parallel speedups for many targets hit by our parallel strategies (see Section \ref{subsec:overall-speedups} and the  lower average speedups for dimension 1000 in Table \ref{table:KdisKrep}).
However, past the last ECD value 
reached by sequential IPOP-CMA-ES, the slope of the curves for the two 
parallel strategies do not bend, showing that these parallel strategies are actually highly efficient in high dimension.}
This is why, along with the parallel speedups, 
ECDF profiles are necessary to appreciate the full extent of the performance of our strategies.
%
%
To finish with, in dimensions 10 and 40 for all granularities, K-Replicated's final runtime reaches slightly more targets than K-Distributed, but only long after K-Distributed \pierre{is over}. 
This is because \pierre{we have let K-Replicated run up to $K_{max}=2^9$ (as opposed to $K_{max}=2^8$ for K-Distributed)} and because K-Replicated's design involves more descents. 
This enables 
K-Replicated to perform more exploration, at the cost of a much greater budget than K-Distributed.
%
}

\tempnewpage

\subsection{Impact of the population size}
\label{section:diffk}

\begin{figure}[t] 
\includegraphics[page=1,width=1\linewidth,trim=0 0 0 0cm,clip]{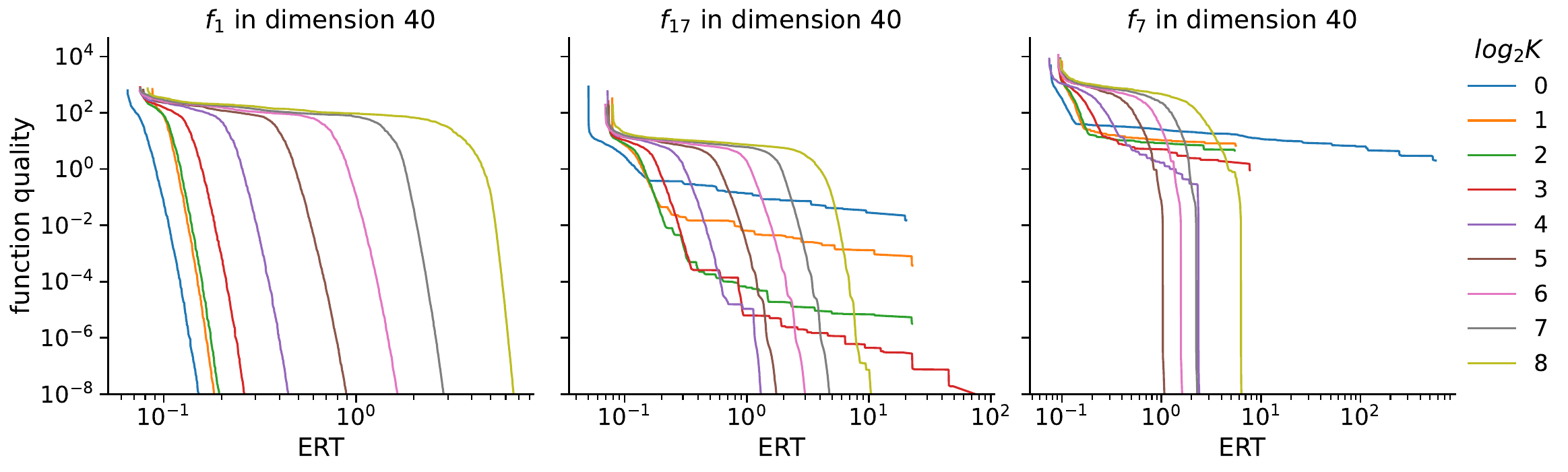}
\caption{
\david{ Function quality over the ERT for the different population sizes of K-Distributed.
}
}
\label{examplePopSize}
\end{figure}

\newcommand{\tableBestK}{\resizebox{.5\columnwidth}{!}{
\begin{tabular}{ c | c|c|c|c|c|c|c|c|c }
& \multicolumn{9}{c}{targets} \\
function & \multicolumn{1}{c}{$10^{2}$} & \multicolumn{1}{c}{$10^{1.5}$} & \multicolumn{1}{c}{$10^{1}$} & \multicolumn{1}{c}{$10^{0.5}$} & \multicolumn{1}{c}{$10^{0}$} & \multicolumn{1}{c}{$10^{-2}$} & \multicolumn{1}{c}{$10^{-4}$} & \multicolumn{1}{c}{$10^{-6}$} & \multicolumn{1}{c}{$10^{-8}$} \\
 \hline
1 & 0.1 & 0.1 & 0.1 & 0.1 & 0.1 & 0.1 & 0.1 & 0.1 & 0.1 \\
2 & 2.2 & 2.4 & 2.4 & 2.7 & 2.8 & 3.0 & 2.8 & 2.8 & 2.7 \\
3 & 0.7 & 3.0 & - & - & - & - & - & - & - \\
4 & 1.3 & 5.1 & - & - & - & - & - & - & - \\
5 & 0.1 & 0.1 & 0.1 & 0.0 & 0.1 & 0.1 & 0.1 & 0.1 & 0.1 \\
6 & 0.1 & 0.1 & 0.1 & 0.3 & 0.4 & 1.1 & 1.0 & 1.2 & 1.3 \\
7 & 0.3 & 0.6 & 1.9 & 3.9 & 4.5 & 4.8 & 4.8 & 4.8 & 4.8 \\
8 & 0.3 & 0.7 & 1.4 & 1.6 & 1.6 & 1.7 & 1.8 & 1.8 & 1.9 \\
9 & 0.3 & 0.8 & 1.7 & 1.9 & 1.9 & 2.0 & 2.0 & 1.9 & 1.9 \\
10 & 1.8 & 1.6 & 1.8 & 1.7 & 1.9 & 2.0 & 2.0 & 2.0 & 2.1 \\
11 & 3.0 & 2.9 & 2.9 & 3.0 & 2.8 & 2.6 & 2.6 & 2.5 & 2.6 \\
12 & 0.2 & 0.3 & 0.7 & 0.9 & 1.1 & 1.2 & 1.2 & 1.6 & 1.6 \\
13 & 0.0 & 0.1 & 0.1 & 0.7 & 1.2 & 2.4 & 2.6 & 3.1 & 3.1 \\
14 & 0.1 & 0.0 & 0.0 & 0.0 & 0.0 & 0.1 & 1.5 & 2.4 & 2.6 \\
15 & 0.6 & 2.1 & 4.2 & 5.6 & 6.4 & 6.5 & 6.5 & 6.5 & 6.5 \\
16 & 0.5 & 1.0 & 1.2 & 0.9 & 1.6 & 6.9 & 7.2 & 7.5 & 7.0 \\
17 & 0.0 & 0.0 & 0.0 & 0.0 & 0.3 & 1.5 & 2.4 & 3.5 & 4.0 \\
18 & 0.1 & 0.1 & 0.1 & 0.5 & 1.2 & 3.7 & 5.4 & 5.7 & 5.7 \\
19 & 0.0 & 0.0 & 0.1 & 1.6 & 1.9 & - & - & - & - \\
20 & 0.0 & 0.1 & 0.1 & 0.0 & 6.9 & - & - & - & - \\
21 & 0.6 & 0.1 & 0.1 & 2.0 & 2.5 & 2.3 & 2.3 & 2.3 & 2.3 \\
22 & 0.0 & 0.1 & 1.4 & 1.4 & 1.0 & - & - & - & - \\
23 & 0.2 & 0.2 & 0.2 & 5.0 & 1.9 & 3.3 & - & - & - \\
24 & 1.0 & 4.2 & 5.0 & 5.7 & - & - & - & - & - \\
\end{tabular}
}}

\begin{table}[t] 
\centering
\caption{ $log_2 K$ (averaged over 20 executions) of the first descent to reach a given quality for a K-Distributed run in dimension 40 (no additional cost). 
’‐’ indicates 
that no descent reached the target.}
\label{tableBestK}
\tableBestK
\end{table}

\begin{figure}[t] 
\includegraphics[width=1\linewidth]{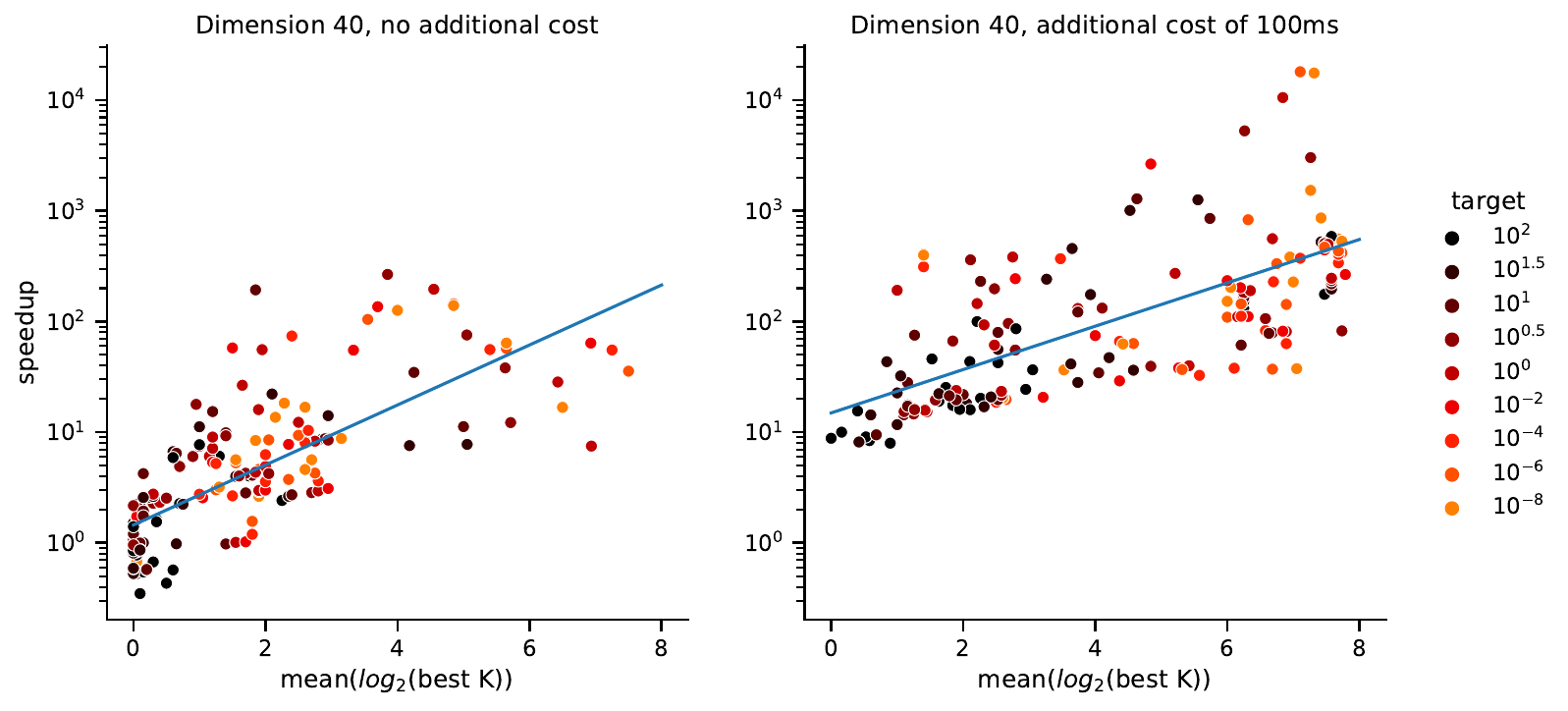}
\caption{Speedup of K-Distributed (over sequential IPOP-CMA-ES) against the best population size for function-target couples,
averaged over 20 executions, for dimension 40, \pierre{with (right) and without (left) additional cost.} 
}
\label{PointsCloudBestK}
\end{figure}

In this section, we analyze the effect of the population size on K-Distributed's performance \pierre{in order to analyze its superior performance.} 
%
\pierre{We start by reporting in} Figure~\ref{examplePopSize} 
the convergence profiles for each distinct population size of K-Distributed on three illustrative BBOB functions.
We observe that the fastest population size to \pierre{reach a function quality} 
varies depending on the targeted function and quality.
For the first \pierre{(i.e. the easiest)} function qualities, \pierre{or for functions with a very simple shape (such as a sphere for $f_1$),}  the descents of K-Distributed are ordered by $K$ value (hence by population size) : the $K=2^0$ 
descent is better than the $K=2^1$ one, 
which is better than the $K=2^2$ one, 
and so forth.
However, for harder function qualities, \pierre{ and for more complex function (such as $f_{17}$),} some descents may stop being competitive after a given \pierre{time}. They typically reach this \pierre{time} 
in the order of $K$ : 
first $2^0$, 
then $2^1$, 
etc. After a descent reaches \pierre{such a time}, 
the descent with the next larger population size 
becomes \pierre{the most} 
time-effective.
We also see that the time before a descent stops \pierre{being effective} can vary greatly with the population size.  
Finally regarding $f_7$, its shape consists in a step
ellipsoidal function which includes many small regions with null gradients. A solver needs good global search abilities to find the optimum, which translates to a large population descent for CMA-ES.
For this function, sequential IPOP-CMA-ES and K-Replicated waste CPU time with the first small population descents 
which last long (especially $K=2^0$) and deliver limited qualities.  
\pierre{This explains in particular the large performance gap between K-Distributed and 
K-Replicated on $f_7$ in Table~\ref{table:KdisKrep}.} 
%
To sum up, this overall dynamic shown in Figure \ref{examplePopSize} makes it difficult to predict which population size will be best for a given problem.

To confirm this analysis, 
we report in Table~\ref{tableBestK} the average $log_2 K$ of the first descent to find a solution for different functions and targets.
We see that lower population sizes are better suited for the first targets (i.e. for the columns with the highest power of 10). 
However, for \pierre{other} 
targets the best population size varies widely, with $log_2 K$ ranging from $0.1$ to $7.5$.
Note that these 
targets are more difficult to solve and \pierre{correspond to} 
the highest speedups of K-Distributed over sequential IPOP-CMA-ES. 
Besides, 
the best population size changes with each function for the final target (column with $10^{-8}$), and also 
with the target for 
each function.
From this, we conclude that no population size is inherently better than another.
Since we lack a reliable way to predict which population size will be more effective, the best strategy is to give an equal chance to each size and start them all at the beginning of the execution, 
which is precisely how K-Distributed operates. 

We want to emphasize that K-Distributed's \pierre{efficiency 
also relies on our parallel evaluations.}
Since \pierre{in our K-Distributed design} the number of cores used is proportional to the population size, the speedups are greater for larger population sizes, which makes 
the duration of the iterations 
closer among different population sizes.
As a result, the convergence of each descent 
operates on a more similar time scale.
Without \pierre{such a parallel evaluation}, 
the descent with larger population sizes would require much more time and would be less competitive 
compared to the ones with lower population sizes.

Lastly, we report in Figure~\ref{PointsCloudBestK} the speedups for K-Distributed over sequential IPOP-CMA-ES 
\pierre{depending on} 
the best population size for each target.
%
\pierre{One can first notice when comparing the two plots that} 
K-Distributed's large populations \pierre{lead to greater speedups} 
when the function evaluation cost is \pierre{longer}. 
This is because \pierre{longer evaluation times 
make the descents less sensitive to the costs of MPI communications and of linear algebra. Descents with large $K$ values can then benefit at best from their higher parallel evaluation speedup.}
%
\pierre{One can also see that} 
K-Distributed's highest speedups are obtained for the larger population sizes.
This is due to sequential IPOP-CMA-ES performing descents in an increasing order of population size, taking more time to start descents of larger populations 
\pierre{(which also applies partly to K-Replicated). 
On the contrary, K-Distributed starts all descents concurrently which is beneficial}
when a large population is \pierre{the most relevant for solving} 
a given problem. 
In some instances, this 
can even produce super-linear speedups as presented in Section \ref{subsec:overall-speedups}.

\section{Conclusion}
\label{section:conclusion}

In this paper, we investigated two parallel strategies for the IPOP-CMA-ES (Covariance Matrix Adaptation Evolution Strategy with Increasing Population) algorithm, 
designed for large blackbox optimization problems \pierre{on thousands of CPU cores}.
Both strategies leveraged BLAS and LAPACK routines to accelerate linear algebra operations, which required 
\pierre{the rewriting of some operations to successfully benefit from the more efficient Level 3 BLAS}. 
The first approach, K‐Replicated, performs multiple descents with identical population sizes, increasing the population size as descents conclude and thus mirroring the progression of IPOP‐CMA‐ES.
On the other hand the second strategy, K‐Distributed, adapts IPOP-CMA-ES 
\pierre{differently} by initiating all descents simultaneously, each with a distinct population size.

\pierre{Thanks to} 
experiments on the supercomputer Fugaku, \pierre{using MPI+OpenMP implementations} on $128$ A64FX CPUs of $48$ cores each ($6144$ cores in total), \pierre{with a reference blackbox optimization benchmark extended with coarser  computation grains}, we determined that these parallel strategies greatly improved on the convergence speed of the original sequential IPOP-CMA-ES, \pierre{reaching speedups up to several thousand}. 
Notably, K‐Distributed outperformed K‐Replicated in \pierre{the vast majority of 
cases}.
Moreover, due to its \pierre{concurrent processing of multiple descents with distinct population sizes,} 
K-Distributed occasionally exhibited super-linear speedups up to $18080\times$ on $6144$ cores.
We complemented these results with a detailed analysis of the superior performance of K-Distributed. 
\pierre{According to our results, if one has a given allocated time on a large-scale parallel architecture, we recommend 
running K-Distributed and possibly restarting each descent once finished until the time is up.}

This study opens up several interesting research avenues.
The large-scale techniques used here could serve as a \pierre{basis} 
for parallelizing other variants of CMA-ES, such as the large-scale 
ones \cite{Rm_CMA_ES,LM_MA_ES,LM_CMA_ES} 
%
Additionally, the parallel IPOP-CMA-ES algorithm could be integrated with other optimization heuristics, such as global optimization \cite{DOOSOO11} or hyper-parameter tuning \cite{CMA_neural}.
Furthermore, investigating methods to predict the most effective CMA-ES population size for an objective function, either beforehand or at runtime, 
\pierre{could strongly benefit to} 
both sequential and parallel blackbox optimization.

\section{Acknowledgments}

Experiments presented in this paper were carried out using the supercomputer Fugaku, provided by the RIKEN Center for Computational Science.
The authors thank the Hauts-de-France region for partly funding David Redon’s PhD thesis.

\bibliography{references}
\end{document}